\documentclass[useAMS,usenatbib]{mn2e}

\usepackage{graphicx}
\usepackage{multirow}

\title[Planetary nebulae near the Galactic centre]{The population of planetary nebulae near the Galactic centre: chemical abundances \thanks{Based on observations obtained at the Southern Astrophysical Research (SOAR) telescope, which is a joint project of the Minist\'erio da Ci\^encia, Tecnologia, e Inova\c{c}\~ao (MCTI) da Rep\'ublica Federativa do Brasil, the U.S. National Optical Astronomy Observatory (NOAO), the University of North Carolina at Chapel Hill (UNC), and Michigan State University (MSU).} \thanks{Baseado em observa\c{c}\~oes realizadas no Observat\'orio do Pico dos Dias / LNA.}}
\author[O. Cavichia, R. D. D. Costa, W. J. Maciel and  M. Moll\'a]{O. Cavichia$^{1,2}$\thanks{E-mail: cavichia@unifei.edu.br}, R. D. D. Costa$^{2}$, W. J. Maciel$^{2}$ and M. Moll\'a$^{3}$\\
$^{1}$Instituto de F\'isica e Qu\'imica, Universidade Federal de Itajub\'a, Av. BPS, 1303, 37500-903, Itajub\'a-MG, Brazil\\
$^{2}$Instituto de Astronomia, Geof\'isica e Ci\^encias Atmosf\'ericas, Universidade de S\~ao Paulo, 05508-900, S\~ao Paulo-SP, Brazil\\
$^{3}$Departamento de Investigaci\'on B\'asica, CIEMAT, Avda. Complutense 40, E-28040 Madrid, Spain}
\begin{document}

\date{Accepted 1988 December 15. Received 1988 December 14; in original form 1988 October 11}

\pagerange{\pageref{firstpage}--\pageref{lastpage}} \pubyear{2002}

\maketitle

\label{firstpage}

\begin{abstract}
Planetary nebulae (PNe) constitute an important tool to study the chemical
evolution of the Milky Way and other galaxies, probing the nucleosynthesis
processes, abundance gradients and the chemical enrichment of the interstellar medium. In particular, Galactic bulge PNe (GBPNe) have been extensively used in the literature to study the chemical properties of this Galactic structure. However, the presently available GBPNe chemical composition studies are strongly biased, since they were focused on brighter objects, predominantly
located in Galactic regions of low interstellar reddening. In this work, we
report physical parameters and abundances derived for a sample of 17
high extinction PNe located in the inner 2\degr of the Galactic bulge,
based on low dispersion spectroscopy secured at the SOAR telescope using the
Goodman spectrograph. The new data allow us to extend our database including
faint objects, providing chemical compositions for PNe located in this region of the bulge and an estimation for the masses of their progenitors to explore the chemical enrichment history of the central region of the Galactic bulge. The results show that there is an enhancement in the N/O  abundance ratio in the Galactic centre PNe compared with PNe located in the outer regions of the Galactic bulge. This may indicate recent episodes of star formation occurring near the Galactic centre.

\end{abstract}


\section{Introduction}

Planetary nebulae (PNe) are the offspring of stars with a large interval of mass ($1-8\, \mbox{M}_\odot$). Since they are formed from several different stellar evolutionary pathways \citep{frew10}, they have very heterogeneous intrinsic and observed properties.  Also, due to their nature, PNe have a very short lifetime dissipating into the interstellar medium (ISM) in a timescale of $3-7 \times10^4$ years \citep{zijlstra91}. PNe constitute an important tool to study the chemical evolution of the Milky Way and other galaxies, probing the nucleosynthesis processes, abundances gradients and the chemical enrichment of the ISM. They provide accurate abundance determinations of several chemical elements difficult to study in stars, such as He and N, and also others such as O, S, Ar, and Ne. The former have abundances modified by the evolution of the PNe progenitor stars, while the latter reflect the conditions of the ISM at the time the progenitors were formed, although a small depletion of O may be observed due to ON cycling for the more massive progenitors \citep{clayton68}, in amounts that depend on metallicity and other properties. Recently, Ne has also been suspected of undergoing self-contamination and its role as a metallicity tracer for the ISM is also uncertain \citep{milingo10}. 

 The currently known number of Galactic bulge planetary nebulae (GBPNe) is $\sim 800$ after the publication of the two Macquire/AAO/Strasbourg H$\alpha$ PNe surveys \citep[MASH][]{parker06,miszalski08}. This number is low compared with the estimated numbers of $\sim 2000$ GBPNe by \citet{gesicki14} and $\sim 3500$ GBPNe by \citet{peyaud06}. The currently number of GBPNe with accurate chemical abundances is $\sim 300$, considering the works done by \citet{escudero01, escudero04, cavichia10}, hereafter IAG-USP sample, and also the works of \citet{gorny09}, \citet{wang07}, \citet{cuisinier00}, \citet{exter04}, as compiled by \citet{chiappini09}. The fact that PNe are originated from stars of different masses and, consequently, evolutionary pathways, hamper the construction of representatives samples for unbiased chemical composition studies. 
 
The presently available chemical composition studies are strongly biased, since they were focused on brighter objects, predominantly located in Galactic regions of low interstellar reddening. The principal obstacle in deriving accurate chemical abundances towards the Galactic centre (GC) is the very high level of extinction close to the Galactic plane, where the interstellar extinction caused by the Galactic disc dust layers can reach $A_V \sim 25 \, \mbox{mag}$ \citep{gonzalez12}. However, it is precisely in the region close to the GC, where a large fraction of the PNe population is expected to exist \citep[][hereafter JS04]{jacoby04}. The chemical abundance distributions of GBPNe in the bulge region within 2$\degr$ of the GC is poorly known, as shown in Fig. 2 of \citet{chiappini09}. The discovery of 160 GBPNe near the GC by JS04 opens the possibility to overcome this bias in the chemical abundance studies. This sample is relatively large, covering nearly two-thirds of the predicted number of GBPNe expected in the surveyed region of the Galactic bulge. Thanks to the access to larger telescopes, now it is possible to secure chemical abundance analysis of this high-extinction and low-surface brightness GBPNe. 

GBPNe near the GC are specially important, since they can shed light on the more general problem of the bulge formation and evolution. For instance, GBPNe can contribute to understand what type of collapse formed the bulge (dissipational or dissipationless), and the role of the secular evolution within the Galaxy. This latter can be probed by GBPNe near the GC, since the secular evolution causes a significant amount of star formation within the centre of galaxies \citep{ellison11}, rejuvenating the stellar populations near the central parsecs of these galaxies \citep{coelho11}. If the Galactic bulge is a classical one, then it is formed by gravitational collapse \citep{eggen62} or by hierarchical merging of smaller objects and the corresponding dissipative gas process \citep{zinn85}. In this case, the formation process is generally fast and occurs earlier in the Galaxy formation process, before the present disc was formed. On the contrary, if the Galactic bulge is formed  by the rearrangement of the disc material, it should be constituted by stellar populations that are similar to those of the inner disc.

A complete survey of the abundances of all known GBPNe is currently beyond our capabilities and we need to observe smaller samples, as in this work, to achieve a more statistically complete coverage of the chemical properties of the intermediate mass population of the bulge. These results would have a significant impact on Galactic evolution theories, providing a much more accurate view of the abundance distribution of GBPNe, especially in the region within 2\degr of the GC and, therefore, producing more reliable constraints for the modelling of intermediate mass stars evolution as well as the chemical evolution of the Galactic disc and bulge. In particular, follow-up spectroscopy of GBPNe can tell us the rate at which the alpha elements were enhanced near the GC. In this paper, for the first time, we report spectroscopic follow-up observations of a sample of GBPNe within a few degrees of the GC. These new data provide chemical composition for PNe located in this region of the bulge and an estimation for the masses of their progenitors to explore the chemical enrichment history of the central region of the Galactic bulge. This paper is organized as follows: Section \ref{sec:obs} describes the criteria for sample selection, the observations and the data reduction. Section \ref{sec:physical_parameters} presents the determination of the physical conditions and the ionic and total abundances of our targets. Section \ref{sec:analysis} is devoted to the data analysis and the comparison of the results obtained in this paper with other results in the literature. Section \ref{sec:conc} summarizes the main findings of this work. 

\section{Sample selection, observations and data reduction}
\label{sec:obs}

\subsection{Sample selection and bulge membership}

The GBPNe population in this work was selected mainly from the catalogue of JS04. In this catalogue, the positions, angular diameters, H$\alpha+$[N\,{\sc ii}] fluxes and 5 GHz fluxes are given for a sample of 94 GBPNe. This survey uses a narrow-band imaging in the near-IR [S\,{\sc iii}] line at 9532~\AA \ to detect highly extinction GBPNe, since this line is the apparently brightest line in the spectra of typical PNe when the $V$-band extinction is between 4 and 12 mag. This point will be addressed in more detail in Section \ref{subsec:data}, where the spectra of the GBPNe of our sample will be presented. JS04 estimated that their survey identified nearly two-thirds of the predicted total number of PNe within $2\degr$ from the GC.  

\begin{figure*}
\includegraphics[width=14.5cm]{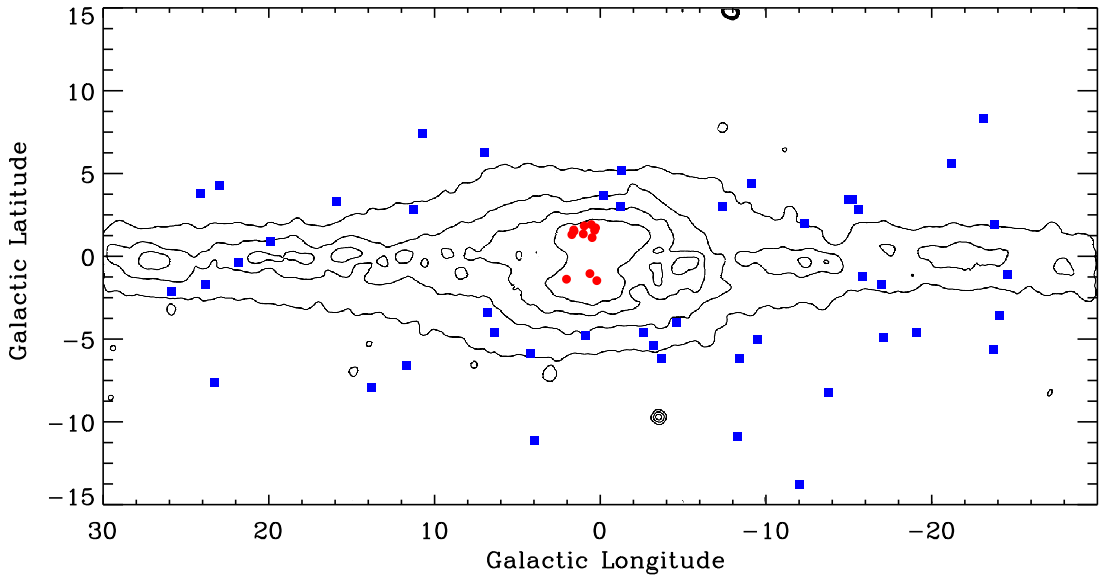}
\caption{The longitude--latitude distribution of the GBPNe from this work (filled red circles) and the data from CCM10 (filled blue squares). The figure also shows the contours of the COBE/DIRBE $2.2 \mu$m image from \citet{weiland94}. Note that only GBPNe from this work with abundances listed in table \ref{tab_elem_abund} are included in this figure.
\label{fig:bulge}}
\end{figure*}

Usually, GBPNe are selected following the standard criteria of \citet{stasinska98}: they have locations within $10\degr$ of the GC, diameters smaller than $12 \arcsec$ and fluxes at 5~GHz lower than 100~mJy. The combination of these criteria leads to the rejection of about $90-95$ per cent of foreground disc PNe that are in the direction of the Galactic bulge. In this paper, these criteria were used whenever possible to select GBPNe from the sample of JS04. Although these criteria may reject some true
extended GBPNe, these objects are very difficult to be detected in the
surveys and, in case of detection, they are extremely faint to provide
accurate fluxes for chemical abundances determinations. Additionally from the sample of JS04, we observed also four objects from the catalogue of \citet{parker06}, that are included in our analysis presented in this work. We performed spectroscopic follow-up of 33 objects located within $2\degr$ of the GC, in a region of a very high-level of reddening. From those objects, 17 had spectra with acceptable
quality to derive physical parameters and chemical abundances. In this paper we present the spectra, physical parameters and chemical abundances for these 17 objects. The longitude-latitude distribution of the sample of GBPNe presented in this work is shown in Fig. \ref{fig:bulge}. In the same figure, the distribution of the data from \citet[][hereafter CCM10]{cavichia10} is shown for comparison. Clearly, the present sample is located in a region much closer to the GC than our previous observations of the Galactic bulge. It is important to note that, to our knowledge, the present sample of GBPNe is the closest sample to the GC than any previous study of chemical composition of GBPNe. 

\begin{figure}
\includegraphics[width=8cm]{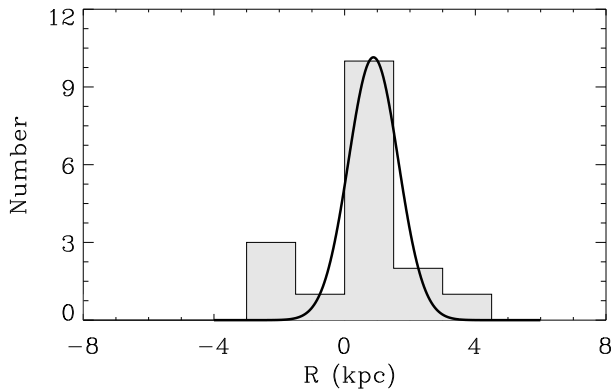}
\caption{Galactocentric distance distribution for the GBPNe from our sample. The continuous line is the result of the histogram data fitted by a Gaussian with mean and standard deviation of $0.89$ kpc and $0.92$ kpc, respectively. 
\label{fig:dist}}
\end{figure}

In order to check the bulge membership of our observed sample, we have calculated individual distances for these objects using the statistical distance scale of \citet{stanghellini08} and the data provided by JS04 for the optical angular diameters and fluxes at 5 GHz. Optical diameters were also taken from \citet{parker06} for the PPA PNe. The statistical distances may present large errors when used individually and in some cases they are as high as 30\% \citep{stanghellini08}. However, when applied to a larger number of objects, they can be very useful to probe the chemical evolution of the Galaxy, as can be seen in \citet{ maciel06, henry10, cavichia11}  and references therein. The logarithmic extinctions at H$\beta$, necessary to derive the distances, were calculated from the fluxes obtained from our own observations. When the fluxes at 5 GHz were not available, equation 6 provided by \citet{cahn92} was used to derive equivalent 5 GHz flux from the H$\beta$ flux. In table \ref{tab:dist}, columns 1 and 2 list the PNG numbers and the PNe names, column 3 the flux in 5 GHz when available or the equivalent flux derived from the H$\beta$ flux. Column 4 of table \ref{tab:dist} shows the optical thickness parameter, as defined by equation 2 of \citet{stanghellini08}, column 5 the optical radius, in arcsec, from JS04 and from \citet{parker06}. The Galactocentric distances obtained from this method are presented in column 6. Recent estimates for the Galactocentric distance of the Sun ($R_0$) ranges from 7.5 to 8.5 kpc, and We have adopted the average value $R_0$ of 8.0 kpc, as suggested by \citet{malkin13}. In table \ref{tab:dist}, the minus sign in the distances indicates the cases when the distances are greater than 8.0 kpc from the Sun. Fig. \ref{fig:dist} shows the galactocentric distance distribution for our sample. The Gaussian fitted to the distribution has mean and standard deviation of 0.89 and 0.92 kpc, respectively. Considering the distance of $R_0$ adopted, the GBPNe studied in this work are $\sim 0.9$ kpc on the average from the GC. Therefore, we are very confident that the present sample is composed by bona fide PNe near the GC. However, for all but one object (JaSt 23) the velocity data are not available in the literature and, since the inner disc can extend into the inner kpc of the Galaxy \citep[see a revision by][]{bland-hawthorn16}, we cannot rule out inner disc PNe contaminants in our sample. In the case of JaSt 23, the OH maser spectrum shows a single peak at $V_{\mbox{\scriptsize LSR}}= +115.2 $ km s$^{-1}$ \citep{uscanga12}, which is compatible with that expected for GBPNe.

 \begin{table}
\footnotesize
  \begin{center}
  \caption{Individual Galactocentric distances  for GBPNe.}
  \label{tab:dist}
  \tabcolsep=0.06cm 
  \begin{tabular}{l l c c c c}\hline \hline
  
  PNG & Name & F  & $\tau$ & $\theta$&  R \\
      &      & (mJy)&      & (arcsec) & (kpc) \\
  \hline
0.344+1.567$\phantom{}^1$ & JaSt 23 &        3 & 4.079 & 3.1 & -2.19 \\
000.2+01.7 & JaSt 19 &        6 & 3.724 & 3.1 & -0.69 \\
000.2-01.4 & JaSt 79 &        4 & 3.785 & 2.5 & -2.69 \\
000.4+01.1 & JaSt 36 &       31 & 2.906 & 2.5 & 0.88 \\
000.5+01.9 & JaSt 17 &       10 & 3.806 & 4.1 & 1.28 \\
000.6-01.0 & JaSt 77 &       49 & 3.309 & 5.1 & 3.72 \\
000.9+01.8 & PPAJ1740-2708  &       23 & 3.192 & 3.1 & 1.25 \\
001.0+01.3 & JaSt 41 &       16 & 3.467 & 3.5 & 1.42 \\
001.5+01.5 & JaSt 46 &       20 & 3.238 & 3.1 & 1.11 \\
001.7+01.3 & JaSt 52 &       24 & 3.017 & 2.5 & 0.54 \\
002.0-01.3 & JaSt 98 &       21 & 3.059 & 2.5 & 0.41 \\
004.3-01.4 & PPAJ1801-2553  &        3 & 3.955 & 2.7 & -2.48 \\
357.7+01.4 & PPAJ1734-3004  &        9 & 3.843 & 4.1 & 1.16 \\
358.5-01.7 & JaSt 64 &       28 & 2.946 & 2.5 & 0.77 \\
358.9-01.5 & JaSt 65 &      122 & 2.308 & 2.5 & 2.59 \\
359.5-01.2 & JaSt 66 &       65 & 2.584 & 2.5 & 1.86 \\
359.9+01.8 & PPAJ1738-2800  &        0 & 4.472 & 2.2 & -9.25 \\
\hline
 \multicolumn{1}{c}{\footnotesize $\phantom{}^1$ OH 0.344 +1.567}						     
  \end{tabular}
 \end{center}
\end{table}
 
\subsection{Observations and data reduction}
\label{subsec:data}

\subsubsection{Optical data}

In 2009 we started an observational program aimed at carrying out a spectroscopic follow-up of GBPNe located within $2 \degr$ of the GC. The 4.1 m SOAR telescope at Cerro Pach\'on (Chile) equipped with the Goodman spectrograph was used for this purpose.  The long-slit spectra were obtained during the years of 2010, 2011 and 2012 using three different VPH gratings 300, 600 and 400 l/mm, respectively. In table \ref{tab:instrum} we list the VPH gratings, slits, wavelength coverage and the FWHM resolution achieved for the optical and near infrared (NIR) observations (see section \ref{nir}). The log of the observations is provided in table \ref{tab:logobs}, as follows: the PNG number (column 1), the name of the object (column 2), the coordinates RA and DEC (columns 3 and 4, respectively), the Galactic longitude and latitude (columns 5 and 6, respectively), the exposition time in seconds (column 7), and the date of observation (column 8). Column 9 lists the VPH grating used in each observation. Columns 10 and 11 list the log of the near infrared observations (see Section \ref{nir}).

\begin{table}
\footnotesize
  \begin{center}
  \caption{Instrumental setup for optical and NIR observations.}
  \label{tab:instrum}
  \begin{tabular}{c c c c}\hline \hline
  
  Grating & Slit width & Coverage  & FWHM  \\
  (l/mm)   & (arcsec) & (\AA)  & (\AA) \\
  \hline
  \multicolumn{4}{c}{Optical}\\
  \hline
  300 & 1.35 & 3600 -- 8800 & 11.8 \\
  400 & 1.68 & 4000 -- 8050 & 11.2\\
  600 & 1.68 & 4700 -- 7360 & 7.3 \\
 \hline
 \multicolumn{4}{c}{NIR}\\
 \hline
 300 & 2.50 & 6010 -- 11010 &  9.0\\
\hline
						     
  \end{tabular}
 \end{center}
\end{table}

The Goodman spectrograph focal plane is imaged onto a Fairchild $4096 \times 4096$ pixels CCD with a read rate of 100 kHz. The exposition times varied between 1800--2600 s, depending on the faintness of the object. Seeing conditions were in general very good ($0.5\arcsec - 0.9\arcsec$). For each night, at least two spectroscopic standard stars were observed, through a $3\arcsec$-wide slit, for flux calibration. Frames containing the same spectral data were combined in order to increase the final signal-to-noise ratio (S/N). The data were reduced with standard procedures by using an IRAF package developed by our group, the PNPACK. This package automatically reduces long-slit spectra by doing bias subtraction, flat field correction, extraction of 1-dimensional spectra, wavelength calibration, atmospheric extinction correction and flux calibration. Cosmic rays were removed using the algorithm for cosmic-ray rejection by Laplacian Edge Detection \citep{vandokkun01}, implemented in the program L.A.Cosmic. In the case of PN spectra, many lines of the observed spectra are weak and, since the objects are located in crowded Galactic bulge fields near the GC, a multi-step method had to be adopted to perform the 1-dimensional spectra extraction. The best-fitting of the background was achieved after removing sky continuum emissions, telluric and interstellar lines, as well as stellar continuum components. The nebular lines were detected and the line fluxes were calculated following an automatic procedure implemented in the PNPACK code. This procedure uses the IRAF task {\it splot} to perform Gaussian fit to emission lines in the spectra and an Gaussian de-blending routine to de-blend lines when necessary. Following this procedure, signal can be attributed to a nebular line only if its value is higher than 2 sigmas of the averaged background noise.

 \begin{table*}
\begin{minipage}{\textwidth}
\footnotesize
  \begin{center}
  \caption{Log of the observations.}
  \label{tab:logobs}
    \tabcolsep=0.11cm 
  \begin{tabular}{c c c c c c c c c  c c}\hline \hline
  
  PNG & Name & RA (J2000) & DEC (J2000) & $\ell (\degr)$ & $b (\degr) $ & $\mbox{T}_{\mbox{exp}}$ (s) & Date & Grating &  $\mbox{T}_{\mbox{exp}}$ (s) $\phantom{}^2$& Date $\phantom{}^2$\\ 
  \hline

0.344+1.567$\phantom{}^1$ & JaSt 23 & 17 40 23.32 &  -27 49 11.7 & 0.35 & 1.57 & 2400 & Jun 28, 11 & 600 & 2400 & Jun 26, 12\\ 
000.2+01.7 & JaSt 19 & 17 39 39.38  &  -27 47 22.58 & 0.28 & 1.72 & 2400 &  Jun 24 , 12 & 300 & 1800 & Jun 26, 14\\
000.2-01.4 & JaSt 79 & 17 51 53.63  &  -29 30 53.41 & 0.21 & -1.47 & 2400 & Jun 08 , 10 & 300 & 2400 & Jun 26, 12 \\
000.4+01.1 & JaSt 36 & 17 42 25.20  &  -27 55 36.36 & 0.49 & 1.13 & 2400 & Jun 12 , 24 & 300 & 2400 & Jun 25, 12\\
000.5+01.9 & JaSt 17 & 17 39 31.22 & -27 27 46.77 & 0.55 & 1.91 & 2400 & Jun 28, 11 & 600 & 2400 & Jun 27, 12\\
000.6-01.0 & JaSt 77 & 17 51 11.65  &  -28 56 27.20 & 0.63 & -1.05 & 1200 & Jun 07 , 10 & 300 & 600 & Jun 26, 14 \\
000.9+01.8 & PPAJ1740-2708 & 17 40 50.70 & -27 08 48.00 & 0.97 & 1.84& 600 &Jun 23, 12 & 400 & -- & -- \\
001.0+01.3 & JaSt 41 & 17 42 49.96  &  -27 21 19.68 & 1.03 & 1.35 & 2400 & Jun 08 , 10 & 300 & 1800 & Jun 25, 12\\
001.5+01.5 & JaSt 46 & 17 43 30.43 & -26 47 32.30 &  1.58 & 1.52 & 2400 & Jun 28, 11 & 600 & 900 & Jun 27, 12\\
001.7+01.3 & JaSt 52 & 17 44 37.30  &  -26 47 25.23 & 1.72 & 1.31 & 1x300 1x2400 & Jun 08 , 10 & 300 & 1200 & Jun 25, 14\\
002.0-01.3 & Jast 98 & 17 55 46.39 & -27 53 38.90 & 2.04 & -1.38 & 2400 & Jun 08, 12 & 400 &  2400 & Jun 26, 12\\
357.7+1.4 &  PPAJ1734-3004  & 17 34 46.6 & -30 04 21 & 357.78 & 1.40 & 2600 & Jun 08, 12 & 400 & -- & -- \\
358.5-01.7 & JaSt 64 & 17 48 56.04  &  -31 06 41.95  & 358.51 & -1.74 & 2x900 & Jun 07 , 10 & 300 & 1800 & Jun 25, 14\\
358.9-01.5 & JaSt 65 & 17 49 20.02  &  -30 36 05.57 & 358.99 & -1.55 & 1800 & Jun 07, 10 & 300 &  2400 & Jun 26, 12\\
359.5-01.2 & JaSt 66 & 17 49 22.10 & -29 59 27.00 & 359.52 &  -1.24 & 2400 & Jun 27 , 11 & 600 & 2400 & Jun 26, 12\\
359.9+01.8 & PPAJ1738-2800 & 17 38 11.80 & -28 00 07.00 & 359.93 & 1.88 & 2400 & Jun 24, 12 & 400 & -- & --\\

 \hline
 \multicolumn{2}{c}{\footnotesize $\phantom{}^1$ OH 0.3447+1.5656}\\		
  \multicolumn{2}{c}{\footnotesize $\phantom{}^2$ Near infrared observations.}	
  \end{tabular}
 \end{center}
 \end{minipage} 
\end{table*}

\subsubsection{Near infrared data \label{nir}}

The optical spectra of the GBPNe near the GC suffer for high-level of extinction caused by the material near the Galactic plane and also in the central regions of the Galaxy. As a result, important diagnostic lines as [O\,{\sc iii}] $\lambda$4363 \AA \ and [N\,{\sc ii}] $\lambda$5755 \AA \ do not have enough S/N ratio to obtain the electron temperature from the temperature diagnostic diagrams. Other important temperature-sensitive lines are those from $S^{+ 3}$. The [S\,{\sc iii}] auroral line at $\lambda$6312 \AA \ is presented in most of our optical spectra. However, the other two [S\,{\sc iii}] lines necessary to obtain the electron temperature are near infrared (NIR) lines at $\lambda \lambda$9069 and 9532 \AA. 

In order to observe the NIR [S\,{\sc iii}] lines, we started an observational program in 2012 at the Observat\'orio Pico dos Dias (OPD) of  National Laboratory for Astrophysics (LNA, Brazil) with the 1.6 m Perkin-Elmer telescope. The spectrophotometry observations were taken according to the log of observations showed in the last two columns of table \ref{tab:logobs}. A Cassegrain Boller \& Chivens spectrograph was used with a 300 l/mm grid, which provides a reciprocal dispersion of 0.22 nm/pixel. An Andor Ikon CCD optimized for NIR observations was used with an operating image scale of 0.56$\arcsec$/pix and a pixel size of 13.5$\times$13.5 $\mu$m. The science spectra were taken with a long slit of 2.5$\arcsec$ with a FWHM spectral resolution of $\sim 9$ \AA. In table \ref{tab:instrum} we also list the setup for the NIR observations. Each night at least two of the spectrophotometric standard stars CD-32 9927, LTT 7379, LTT 9239,  CD-34 241 of \citet{hamuy92, hamuy94} were observed to improve the flux calibration. These stars were observed with a long slit of 7.5$\arcsec$ width, allowing a more precise flux calibration. Helium-argon arcs were taken immediately after each science spectra in order to perform wavelength calibration. Telluric corrections were not performed since the [S\,{\sc iii}] $\lambda$9069 and 9532\AA \ are relatively very intense and telluric absorption effects are negligible for these lines. This point will be addressed again in section \ref{sec:reddening}, were the observed relative fluxes of these lines are confronted with the expected theoretical values. Data reduction was performed using the IRAF package, following the standard procedure for long slit spectra, as in the previous section: correction of bias, flat-field, extraction, wavelength and flux calibration. Atmospheric extinction was corrected through mean coefficients derived for the LNA observatory. 

\begin{figure}
\includegraphics[width=8cm]{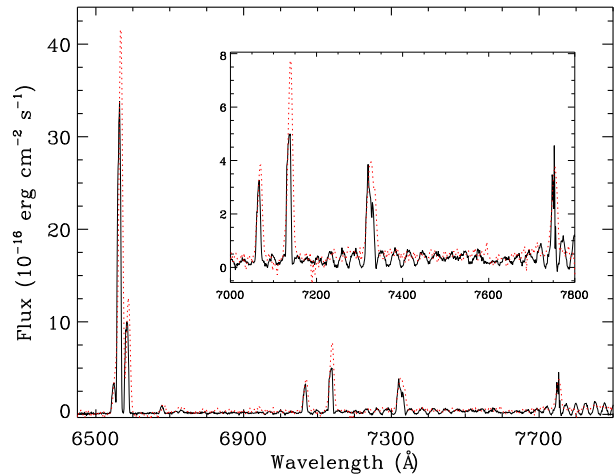}
\caption{Calibrated optical (black continuous line) and NIR (dashed red line) spectra for JaSt 98 in the region where the line-ratios were used to calculate the normalizing constant. At the top-right panel it is shown a zoom in the faintest lines. \label{fig:spec_match}}
\end{figure}

The NIR spectra were normalized to match their optical counterparts using as many as lines as available in the spectra. The normalizing constant for each NIR spectra was obtained by the flux weighted mean of the ratio between the NIR line and the optical counterpart. In general the concordance in flux was good between both spectra (optical and NIR), and the mean ratio of the lines varying between 0.7 and 1.7. The variation of the normalizing constant can be attributed to some causes: atmospheric extinction correction, as we are using the mean extinction absorption coefficients provided by the observatories; the slit widths are different in the optical and NIR observations and this can cause some fluxes loss; cirrus clouds can cause a grey extinction in the spectra and, therefore, change the absolute flux values. In Fig. \ref{fig:spec_match} an example is presented showing the match between the optical and NIR spectra in the region were there are common lines in both observations. It is important to note in this figure that due to the different instrumental setups in both observations, the line profiles are different and the peaks of the lines may not match. The final spectra calibrated in flux and wavelength are shown in Fig. \ref{fig:spectra1}. Note that in this figure the spectra were not corrected for interstellar extinction. The reddened fluxes of the GBPNe are listed in Table \ref{tab_fluxes}. In this table, the fluxes are presented relative to the one from H$\beta$ line. The errors associated with the fluxes were attributed based on the intensity of the line, as will be explained in Section \ref{sec:errors}.

\begin{figure*}
\includegraphics[width=16cm]{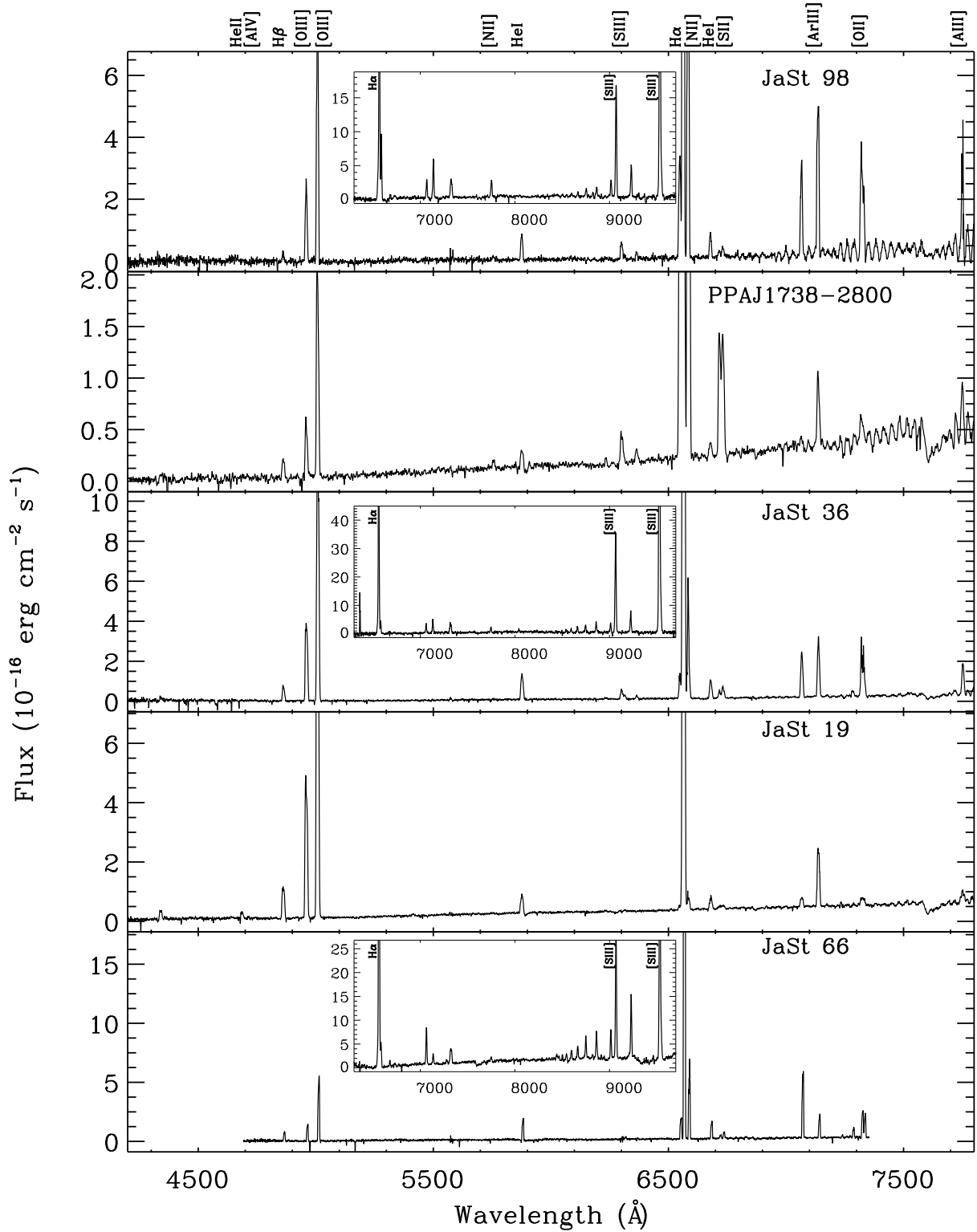}
\caption{Calibrated optical spectra of low and medium resolution of the GBPNe observed. At the center-top of each optical spectra it is shown the respective NIR spectrum, when available.\label{fig:spectra1}}
\end{figure*}

\setcounter{figure}{3} 

\begin{figure*}
\includegraphics[width=16cm]{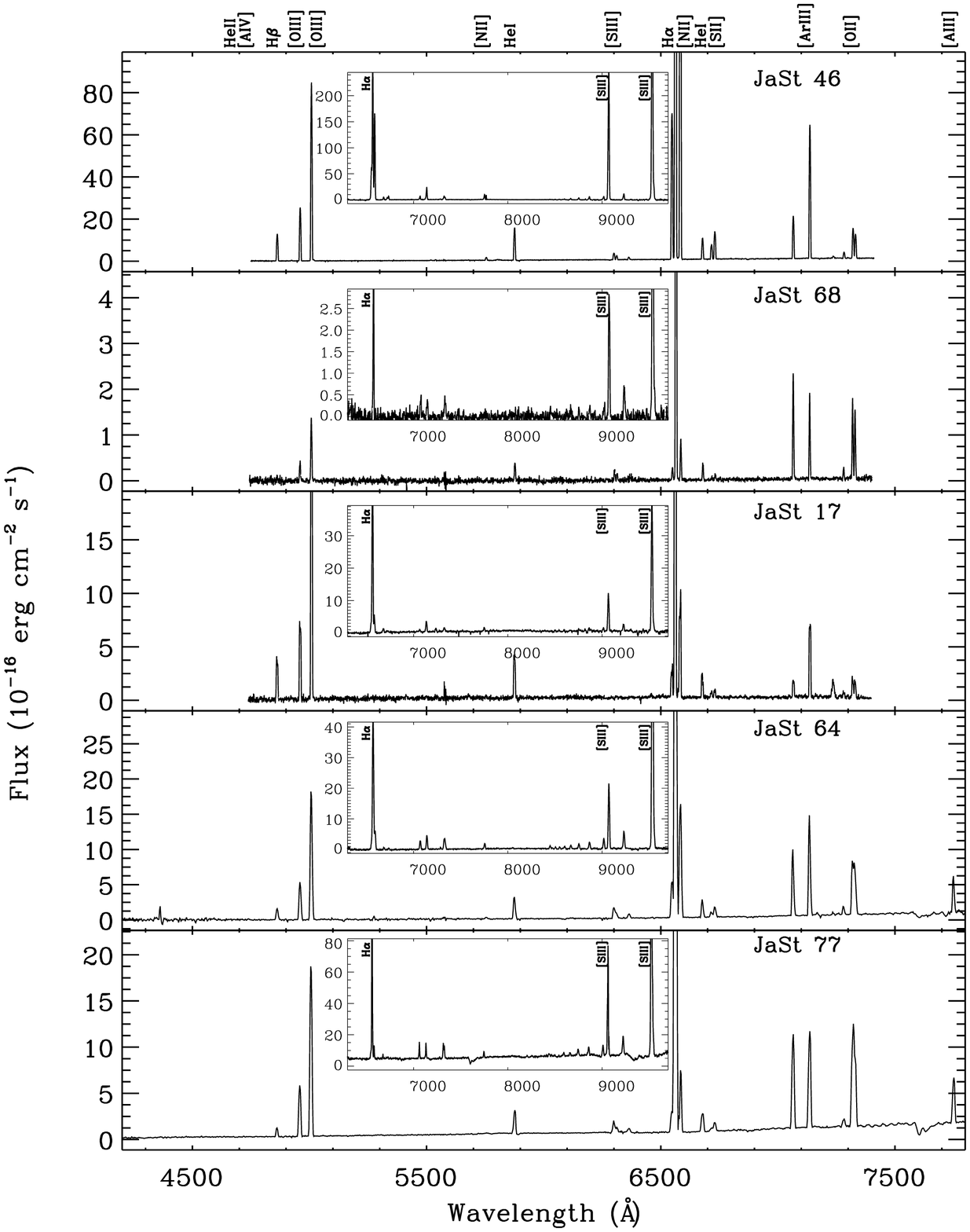}
\caption{Continued.}
\end{figure*}

\setcounter{figure}{3} 

\begin{figure*}
\includegraphics[width=16cm]{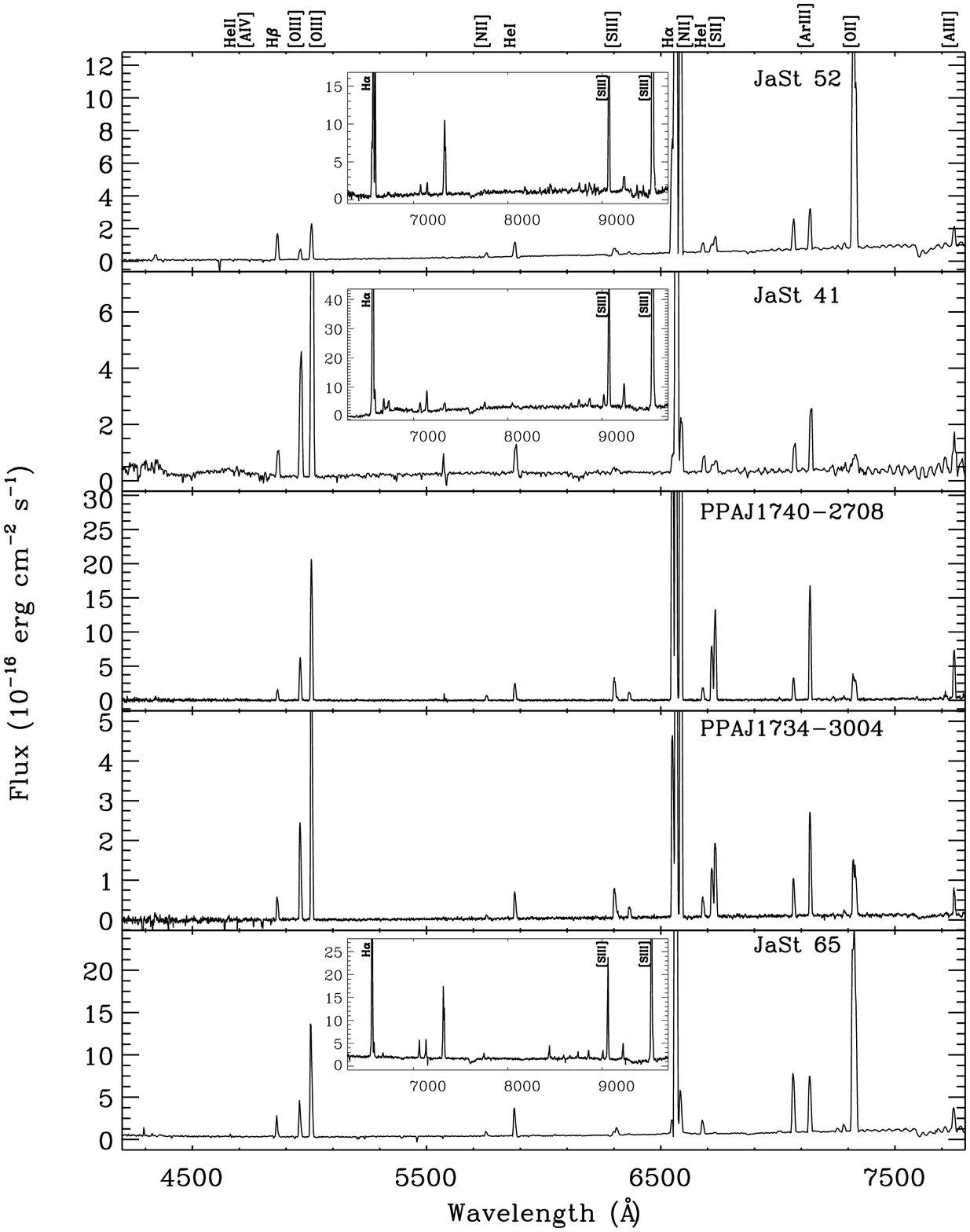}
\caption{Continued.}
\end{figure*}

\setcounter{figure}{3} 

\begin{figure*} 
\includegraphics[width=16cm]{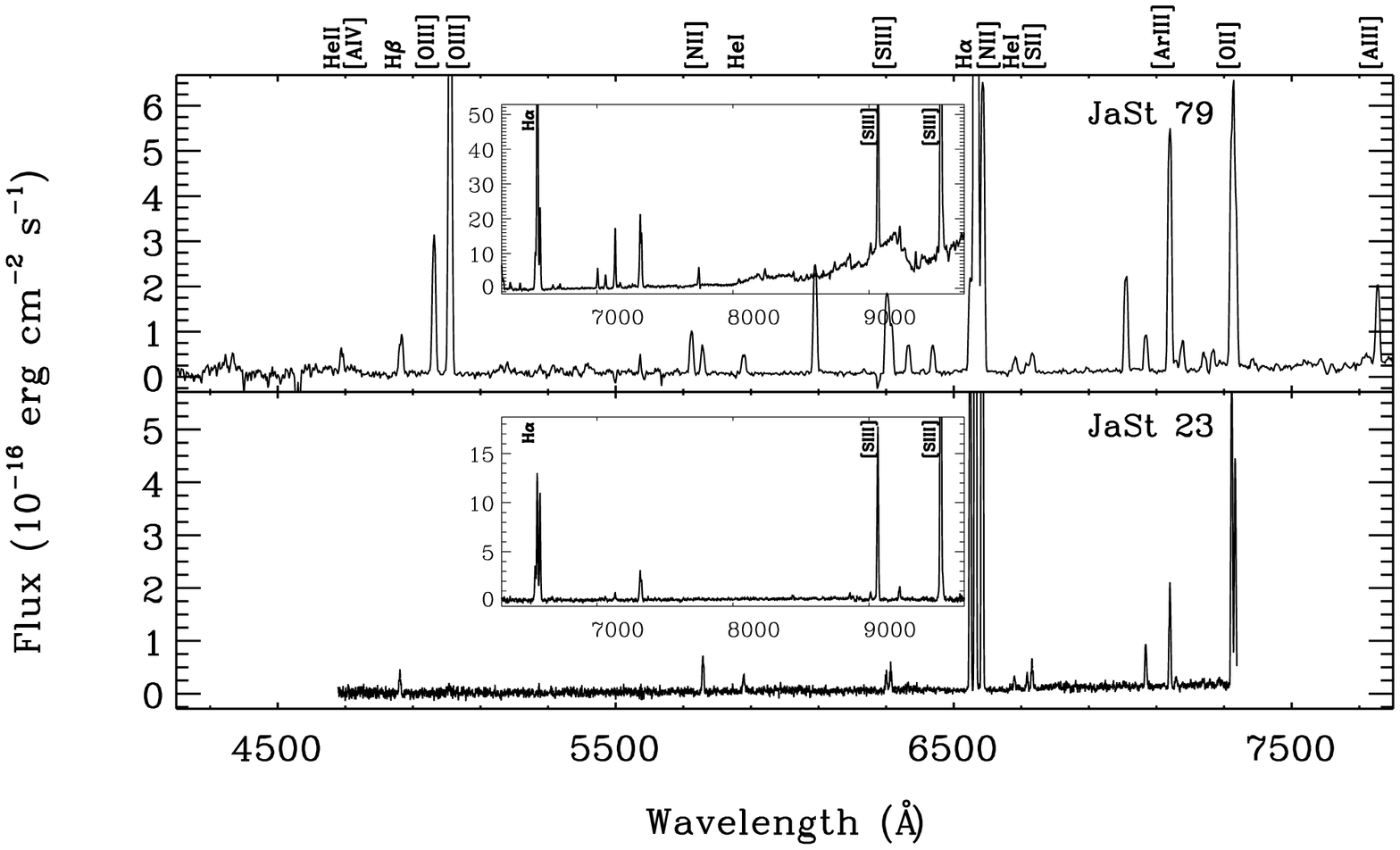}
\caption{Continued.}
\end{figure*}

\begin{table*}
  \begin{center}
  \caption{Reddened fluxes relative to H$\beta$. $F(\mbox{H}\beta)$ is in units of erg cm$^{-2}$ s$^{-1}$. }
  \label{tab_fluxes}
  \begin{tabular}{l r c c c c c c c c c c c c c c}
  
  \hline \hline

&  & \multicolumn{2}{c}{0.344+1.567$\phantom{}^1$}   & \multicolumn{2}{c}{000.2+01.7}   & \multicolumn{2}{c}{000.2-01.4}   & \multicolumn{2}{c}{000.4+01.1}   \\
&  & \multicolumn{2}{c}{JaSt 23}   & \multicolumn{2}{c}{JaSt 19}   & \multicolumn{2}{c}{JaSt 79}   & \multicolumn{2}{c}{JaSt 36}   \\
$\lambda$ & Ion & $F(\lambda)/F(\mbox{H}\beta)$ & Error & $F(\lambda)/F(\mbox{H}\beta)$ & Error & $F(\lambda)/F(\mbox{H}\beta)$ & Error & $F(\lambda)/F(\mbox{H}\beta)$ & Error \\
\hline
 4341. & H$\gamma$ & -- & -- &       0.24 &       0.02 &       0.32 &       0.03 &       0.17 &       0.02 \\
 4363. & OIII & -- & -- &       0.04 &       0.01 &       0.39 &       0.04 & -- & -- \\
 4686. & HeII & -- & -- &       0.15 &       0.02 &       0.46 &       0.05 & -- & -- \\
 4861. & H$\beta$+HeII &       1.00 &       0.10 &       1.00 &       0.10 &       1.00 &       0.10 &       1.00 &       0.10 \\
 4959. & OIII & -- & -- &       3.95 &       0.40 &       3.17 &       0.32 &       5.65 &       0.28 \\
 5007. & OIII &       0.22 &       0.02 &      13.29 &       0.66 &      10.82 &       0.54 &      17.70 &       0.89 \\
 5412. & HeII & -- & -- &       0.06 &       0.01 &       0.38 &       0.04 & -- & -- \\
 5518. & Cl3 & -- & -- & -- & -- & -- & -- & -- & -- \\
 5538. & Cl3 & -- & -- & -- & -- & -- & -- & -- & -- \\
 5755. & NII &       1.66 &       0.17 &       0.02 &       0.01 &       0.60 &       0.06 &       0.03 &       0.01 \\
 5876. & HeI &       0.85 &       0.09 &       0.52 &       0.05 &       0.51 &       0.05 &       1.76 &       0.18 \\
 6300. & OI &       0.79 &       0.08 & -- & -- &       2.06 &       0.21 &       0.54 &       0.05 \\
 6312. & SIII &       1.09 &       0.11 &       0.03 &       0.01 &       1.11 &       0.11 &       0.29 &       0.03 \\
 6363. & OI & -- & -- & -- & -- &       0.71 &       0.07 &       0.19 &       0.02 \\
 6435. & ArV & -- & -- & -- & -- &       0.66 &       0.07 & -- & -- \\
 6548. & NII &      19.18 &       0.96 &       0.11 &       0.01 &       2.21 &       0.22 &       1.56 &       0.16 \\
 6563. & H$\alpha$ &      73.86 &       3.69 &      30.06 &       1.50 &      36.68 &       1.83 &      74.44 &       3.72 \\
 6584. & NII &      56.23 &       2.81 &       0.35 &       0.03 &       6.64 &       0.33 &       4.91 &       0.49 \\
 6678. & HeI &       0.85 &       0.09 &       0.38 &       0.04 &       0.44 &       0.04 &       1.26 &       0.13 \\
 6716. & SII &       0.76 &       0.08 &       0.10 &       0.02 &       0.23 &       0.02 &       0.47 &       0.05 \\
 6731. & SII &       1.42 &       0.14 &       0.12 &       0.01 &       0.51 &       0.05 &       0.77 &       0.08 \\
 7005. & ArV & -- & -- & -- & -- &       2.37 &       0.24 & -- & -- \\
 7065. & HeI &       2.19 &       0.22 &       0.30 &       0.03 &       0.89 &       0.09 &       2.91 &       0.29 \\
 7135. & ArIII &       4.66 &       0.47 &       1.66 &       0.17 &       6.05 &       0.30 &       3.59 &       0.36 \\
 7237. & ArIV & -- & -- &       0.07 &       0.01 &       0.44 &       0.04 & -- & -- \\
 7263. & ArIV & -- & -- & -- & -- &       0.50 &       0.05 & -- & -- \\
 7281. & HeI & -- & -- & -- & -- & -- & -- &       0.41 &       0.04 \\
 7322. & OII &      14.79 &       0.74 &       0.37 &       0.04 &       9.76 &       0.49 &       2.28 &       0.23 \\
 7333. & OII &      10.26 &       0.51 & -- & -- & -- & -- &       1.98 &       0.20 \\
 7751. & ArIII & -- & -- &       0.44 &       0.04 &       1.98 &       0.20 &       1.91 &       0.19 \\
 9069. & SIII &     115.23 &       5.76 & -- & -- &      21.11 &       1.06 &      49.06 &       2.45 \\
 9532. & SIII &     355.35 &      17.77 & -- & -- &      52.47 &       2.62 &     151.65 &       7.58 \\
 $\log(F\mbox{H}\beta)$ &  &    -15.602 &   &    -14.856 &   &    -14.919 &   &    -15.079 &   \\

 \hline						     
  \end{tabular}
 \end{center}
\end{table*}

\setcounter{table}{3} 

\begin{table*}
  \begin{center}
  \caption{Continued.}
  \label{tab_fluxes}
  \begin{tabular}{l r c c c c c c c c c c c c c c}
  
  \hline \hline
&  & \multicolumn{2}{c}{000.5+01.9}   & \multicolumn{2}{c}{000.6-01.0}   & \multicolumn{2}{c}{000.9+01.8}   & \multicolumn{2}{c}{001.0+01.3}   \\
&  & \multicolumn{2}{c}{JaSt 17}   & \multicolumn{2}{c}{JaSt 77}   & \multicolumn{2}{c}{PPAJ1740-2708 }   & \multicolumn{2}{c}{JaSt 41}   \\
$\lambda$ & Ion & $F(\lambda)/F(\mbox{H}\beta)$ & Error & $F(\lambda)/F(\mbox{H}\beta)$ & Error & $F(\lambda)/F(\mbox{H}\beta)$ & Error & $F(\lambda)/F(\mbox{H}\beta)$ & Error \\
\hline
 4341. & H$\gamma$ & -- & -- & -- & -- &       0.27 &       0.03 & -- & -- \\
 4363. & OIII & -- & -- & -- & -- & -- & -- & -- & -- \\
 4686. & HeII & -- & -- & -- & -- &       0.16 &       0.02 & -- & -- \\
 4861. & H$\beta$+HeII &       1.00 &       0.10 &       1.00 &       0.10 &       1.00 &       0.10 &       1.00 &       0.10 \\
 4959. & OIII &       1.85 &       0.19 &       5.79 &       0.29 &       4.01 &       0.40 &       1.99 &       0.20 \\
 5007. & OIII &       6.30 &       0.32 &      19.85 &       0.99 &      13.46 &       0.67 &       6.62 &       0.33 \\
 5412. & HeII &       0.06 &       0.01 & -- & -- &       0.05 &       0.01 & -- & -- \\
 5518. & Cl3 & -- & -- & -- & -- & -- & -- & -- & -- \\
 5538. & Cl3 & -- & -- & -- & -- & -- & -- & -- & -- \\
 5755. & NII &       0.03 &       0.01 & -- & -- &       0.51 &       0.05 & -- & -- \\
 5876. & HeI &       1.01 &       0.10 &       2.44 &       0.24 &       1.72 &       0.17 &       2.01 &       0.20 \\
 6300. & OI & -- & -- &       1.15 &       0.11 &       2.29 &       0.23 &       0.38 &       0.04 \\
 6312. & SIII &       0.04 &       0.01 &       0.63 &       0.06 &       0.39 &       0.04 &       0.28 &       0.03 \\
 6363. & OI & -- & -- &       0.58 &       0.06 &       0.89 &       0.09 & -- & -- \\
 6435. & ArV & -- & -- & -- & -- & -- & -- & -- & -- \\
 6548. & NII &       0.80 &       0.08 &       1.92 &       0.19 &      32.10 &       1.61 &       2.12 &       0.21 \\
 6563. & H$\alpha$ &      24.91 &       1.25 &     131.13 &       6.56 &      48.23 &       2.41 &      86.62 &       4.33 \\
 6584. & NII &       2.51 &       0.25 &       5.76 &       0.29 &     105.26 &       5.26 &       6.37 &       0.32 \\
 6678. & HeI &       0.52 &       0.05 &       2.82 &       0.28 &       1.36 &       0.14 &       1.66 &       0.17 \\
 6716. & SII &       0.13 &       0.01 &       0.53 &       0.05 &       5.56 &       0.28 &       0.34 &       0.03 \\
 6731. & SII &       0.18 &       0.02 &       1.10 &       0.11 &       9.28 &       0.46 &       0.50 &       0.05 \\
 7005. & ArV & -- & -- & -- & -- &       0.25 &       0.02 & -- & -- \\
 7065. & HeI &       0.42 &       0.04 &      11.67 &       0.58 &       2.25 &       0.23 &       6.53 &       0.33 \\
 7135. & ArIII &       1.81 &       0.18 &      12.30 &       0.62 &      10.98 &       0.55 &       2.35 &       0.23 \\
 7237. & ArIV &       0.55 &       0.06 & -- & -- &       0.38 &       0.04 & -- & -- \\
 7263. & ArIV & -- & -- & -- & -- & -- & -- & -- & -- \\
 7281. & HeI &       0.14 &       0.01 &       1.07 &       0.11 &       0.25 &       0.03 &       0.93 &       0.09 \\
 7322. & OII &       0.41 &       0.04 &      19.55 &       0.98 &       4.43 &       0.44 &       4.51 &       0.45 \\
 7333. & OII &       0.38 &       0.04 & -- & -- & -- & -- & -- & -- \\
 7751. & ArIII & -- & -- &       5.69 &       0.28 &       4.79 &       0.48 &       1.25 &       0.13 \\
 9069. & SIII & -- & -- &      97.83 &       4.89 & -- & -- &      37.53 &       1.88 \\
 9532. & SIII & -- & -- &     303.73 &      15.19 & -- & -- &     114.70 &       5.74 \\
 $\log(F\mbox{H}\beta)$ &  &    -14.507 &   &    -14.995 &   &    -14.852 &   &    -15.422 &   \\

 \hline						     
  \end{tabular}
 \end{center}
\end{table*}

\setcounter{table}{3} 

\begin{table*}
  \begin{center}
  \caption{Continued.}
  \label{tab_fluxes}
  \begin{tabular}{l r c c c c c c c c c c c c c c}
  
  \hline \hline

&  & \multicolumn{2}{c}{001.5+01.5}   & \multicolumn{2}{c}{001.6+01.5}   & \multicolumn{2}{c}{001.7+01.3}   & \multicolumn{2}{c}{002.0-01.3}   \\
&  & \multicolumn{2}{c}{JaSt 46}   & \multicolumn{2}{c}{JaSt 42}   & \multicolumn{2}{c}{JaSt 52}   & \multicolumn{2}{c}{JaSt 98}   \\
$\lambda$ & Ion & $F(\lambda)/F(\mbox{H}\beta)$ & Error & $F(\lambda)/F(\mbox{H}\beta)$ & Error & $F(\lambda)/F(\mbox{H}\beta)$ & Error & $F(\lambda)/F(\mbox{H}\beta)$ & Error \\
\hline
 4341. & H$\gamma$ & -- & -- & -- & -- & -- & -- & -- & -- \\
 4363. & OIII & -- & -- & -- & -- & -- & -- & -- & -- \\
 4686. & HeII & -- & -- & -- & -- & -- & -- & -- & -- \\
 4861. & H$\beta$+HeII &       1.00 &       0.10 &       1.00 &       0.10 &       1.00 &       0.10 &       1.00 &       0.10 \\
 4959. & OIII &       1.92 &       0.19 &       2.76 &       0.28 &       0.40 &       0.04 &       8.76 &       0.44 \\
 5007. & OIII &       6.27 &       0.31 &       9.32 &       0.47 &       1.35 &       0.13 &      27.15 &       1.36 \\
 5412. & HeII & -- & -- & -- & -- & -- & -- & -- & -- \\
 5518. & Cl3 &       0.03 &       0.01 & -- & -- & -- & -- & -- & -- \\
 5538. & Cl3 &       0.03 &       0.01 & -- & -- & -- & -- & -- & -- \\
 5755. & NII &       0.10 &       0.01 & -- & -- &       0.18 &       0.02 &       0.25 &       0.03 \\
 5876. & HeI &       1.16 &       0.12 &       0.66 &       0.07 &       0.57 &       0.06 &       3.03 &       0.30 \\
 6300. & OI &       0.25 &       0.02 &       0.02 &       0.00 &       0.34 &       0.03 &       2.20 &       0.22 \\
 6312. & SIII &       0.14 &       0.01 &       0.09 &       0.02 &       0.07 &       0.01 &       0.42 &       0.04 \\
 6363. & OI &       0.09 &       0.02 & -- & -- &       0.08 &       0.02 &       0.72 &       0.07 \\
 6435. & ArV & -- & -- & -- & -- & -- & -- &       0.30 &       0.03 \\
 6548. & NII &       5.46 &       0.27 &       0.77 &       0.08 &       4.62 &       0.46 &      11.94 &       0.60 \\
 6563. & H$\alpha$ &      35.38 &       1.77 &      42.82 &       2.14 &      42.84 &       2.14 &     126.57 &       6.33 \\
 6584. & NII &      17.92 &       0.90 &       2.39 &       0.24 &      14.78 &       0.74 &      37.91 &       1.90 \\
 6678. & HeI &       0.76 &       0.08 &       0.81 &       0.08 &       0.40 &       0.04 &       2.74 &       0.27 \\
 6716. & SII &       0.55 &       0.05 &       0.23 &       0.02 &       0.34 &       0.03 &       0.84 &       0.08 \\
 6731. & SII &       1.03 &       0.10 &       0.34 &       0.03 &       0.67 &       0.07 &       1.29 &       0.13 \\
 7005. & ArV & -- & -- & -- & -- &       0.14 &       0.01 & -- & -- \\
 7065. & HeI &       1.45 &       0.15 &       1.03 &       0.10 &       1.27 &       0.13 &      10.78 &       0.54 \\
 7135. & ArIII &       4.53 &       0.45 &       3.63 &       0.36 &       1.68 &       0.17 &      19.65 &       0.98 \\
 7237. & ArIV &       0.12 &       0.01 &       0.50 &       0.05 & -- & -- & -- & -- \\
 7263. & ArIV & -- & -- & -- & -- & -- & -- & -- & -- \\
 7281. & HeI &       0.21 &       0.02 & -- & -- & -- & -- & -- & -- \\
 7322. & OII &       1.04 &       0.10 &       0.86 &       0.09 &      15.53 &       0.78 &      18.24 &       0.91 \\
 7333. & OII &       0.84 &       0.08 & -- & -- & -- & -- & -- & -- \\
 7751. & ArIII & -- & -- &       2.46 &       0.25 &       0.78 &       0.08 &      11.36 &       0.57 \\
 9069. & SIII & -- & -- &      31.80 &       1.59 & -- & -- &      79.85 &       3.99 \\
 9532. & SIII & -- & -- &      92.42 &       4.62 & -- & -- &     249.31 &      12.47 \\
 $\log(F\mbox{H}\beta)$ &  &    -14.048 &   &    -14.864 &   &    -14.745 &   &    -15.572 &   \\

 \hline						     
  \end{tabular}
 \end{center}
\end{table*}

\setcounter{table}{3} 

\begin{table*}
  \begin{center}
  \caption{Continued.}
  \label{tab_fluxes}
  \begin{tabular}{l r c c c c c c c c c c c c c c}
  
  \hline \hline

&  & \multicolumn{2}{c}{357.7+01.4}   & \multicolumn{2}{c}{358.5-01.7}   & \multicolumn{2}{c}{358.9-01.5}   & \multicolumn{2}{c}{359.5-01.2}   \\
&  & \multicolumn{2}{c}{PPAJ1734-3004 }   & \multicolumn{2}{c}{JaSt 64}   & \multicolumn{2}{c}{JaSt 65}   & \multicolumn{2}{c}{JaSt 66}   \\
$\lambda$ & Ion & $F(\lambda)/F(\mbox{H}\beta)$ & Error & $F(\lambda)/F(\mbox{H}\beta)$ & Error & $F(\lambda)/F(\mbox{H}\beta)$ & Error & $F(\lambda)/F(\mbox{H}\beta)$ & Error \\
\hline
 4341. & H$\gamma$ &       0.27 &       0.03 & -- & -- & -- & -- & -- & -- \\
 4363. & OIII & -- & -- & -- & -- & -- & -- & -- & -- \\
 4686. & HeII & -- & -- & -- & -- & -- & -- & -- & -- \\
 4861. & H$\beta$+HeII &       1.00 &       0.10 &       1.00 &       0.10 &       1.00 &       0.10 &       1.00 &       0.10 \\
 4959. & OIII &       4.27 &       0.43 &       3.44 &       0.34 &       2.04 &       0.20 &       1.79 &       0.18 \\
 5007. & OIII &      14.35 &       0.72 &      11.34 &       0.57 &       6.97 &       0.35 &       6.50 &       0.33 \\
 5412. & HeII & -- & -- & -- & -- & -- & -- & -- & -- \\
 5518. & Cl3 & -- & -- & -- & -- & -- & -- & -- & -- \\
 5538. & Cl3 & -- & -- & -- & -- & -- & -- & -- & -- \\
 5755. & NII &       0.20 &       0.02 &       0.14 &       0.01 &       0.35 &       0.03 & -- & -- \\
 5876. & HeI &       1.23 &       0.12 &       1.84 &       0.18 &       1.83 &       0.18 &       2.16 &       0.22 \\
 6300. & OI &       1.27 &       0.13 &       0.85 &       0.09 &       0.20 &       0.02 & -- & -- \\
 6312. & SIII &       0.63 &       0.06 &       0.57 &       0.06 &       0.67 &       0.07 &       0.26 &       0.03 \\
 6363. & OI &       0.60 &       0.06 &       0.41 &       0.04 & -- & -- &       0.10 &       0.01 \\
 6435. & ArV & -- & -- & -- & -- & -- & -- & -- & -- \\
 6548. & NII &       9.45 &       0.47 &       3.71 &       0.37 &       1.26 &       0.13 &       2.41 &       0.24 \\
 6563. & H$\alpha$ &      49.77 &       2.49 &      84.12 &       4.21 &      77.70 &       3.89 &     105.63 &       5.28 \\
 6584. & NII &      29.32 &       1.47 &      11.63 &       0.58 &       3.77 &       0.38 &       7.49 &       0.37 \\
 6678. & HeI &       0.97 &       0.10 &       1.39 &       0.14 &       0.98 &       0.10 &       1.75 &       0.18 \\
 6716. & SII &       2.41 &       0.24 &       0.47 &       0.05 & -- & -- &       0.43 &       0.04 \\
 6731. & SII &       3.88 &       0.39 &       0.95 &       0.09 &       0.10 &       0.02 &       0.64 &       0.06 \\
 7005. & ArV & -- & -- & -- & -- &       0.24 &       0.02 & -- & -- \\
 7065. & HeI &       1.72 &       0.17 &       5.17 &       0.26 &       4.72 &       0.47 &       6.68 &       0.33 \\
 7135. & ArIII &       4.71 &       0.47 &       8.07 &       0.40 &       4.52 &       0.45 &       2.37 &       0.24 \\
 7237. & ArIV & -- & -- &       0.20 &       0.02 & -- & -- &       0.18 &       0.02 \\
 7263. & ArIV & -- & -- & -- & -- & -- & -- & -- & -- \\
 7281. & HeI &       0.33 &       0.03 &       0.66 &       0.07 &       0.58 &       0.06 &       1.00 &       0.10 \\
 7322. & OII &       2.67 &       0.27 &       8.94 &       0.45 &      25.68 &       1.28 &       2.87 &       0.29 \\
 7333. & OII &       2.06 &       0.21 & -- & -- & -- & -- &       2.33 &       0.23 \\
 7751. & ArIII &       1.25 &       0.12 &       3.04 &       0.30 &       1.83 &       0.18 & -- & -- \\
 9069. & SIII & -- & -- &      80.96 &       4.05 &      29.40 &       1.47 &      36.63 &       1.83 \\
 9532. & SIII & -- & -- &     245.67 &      12.28 &      88.57 &       4.43 &     112.91 &       5.65 \\
 $\log(F\mbox{H}\beta)$ &  &    -15.294 &   &    -14.791 &   &    -14.746 &   &    -15.225 &   \\

 \hline						     
  \end{tabular}
 \end{center}
\end{table*}

\setcounter{table}{3} 

\begin{table*}
  \begin{center}
  \caption{Continued.}
  \label{tab_fluxes}
  \begin{tabular}{l r c c c c c c c c c c c c c c}
  
  \hline \hline

&  & \multicolumn{2}{c}{359.5-01.3}   & \multicolumn{2}{c}{359.9+01.8}           \\
&  & \multicolumn{2}{c}{JaSt 68}   & \multicolumn{2}{c}{PPAJ1738-2800 }           \\
$\lambda$ & Ion & $F(\lambda)/F(\mbox{H}\beta)$ & Error & $F(\lambda)/F(\mbox{H}\beta)$ & Error         \\
\hline
 4341. & H$\gamma$ & -- & -- & -- & --         \\
 4363. & OIII & -- & -- & -- & --         \\
 4686. & HeII & -- & -- & -- & --         \\
 4861. & H$\beta$+HeII &       1.00 &       0.10 &       1.00 &       0.10         \\
 4959. & OIII &       3.76 &       0.38 &       2.23 &       0.22         \\
 5007. & OIII &      13.09 &       0.65 &       7.95 &       0.40         \\
 5412. & HeII & -- & -- & -- & --         \\
 5518. & Cl3 & -- & -- & -- & --         \\
 5538. & Cl3 & -- & -- & -- & --         \\
 5755. & NII & -- & -- &       0.37 &       0.04         \\
 5876. & HeI &       3.77 &       0.38 &       0.70 &       0.07         \\
 6300. & OI &       2.06 &       0.21 &       1.14 &       0.11         \\
 6312. & SIII &       1.40 &       0.14 &       0.17 &       0.02         \\
 6363. & OI & -- & -- &       0.55 &       0.06         \\
 6435. & ArV & -- & -- & -- & --         \\
 6548. & NII &       2.68 &       0.27 &      10.76 &       0.54         \\
 6563. & H$\alpha$ &     192.20 &       9.61 &      22.49 &       1.12         \\
 6584. & NII &       7.72 &       0.39 &      33.70 &       1.69         \\
 6678. & HeI &       3.29 &       0.33 &       0.61 &       0.06         \\
 6716. & SII &       0.57 &       0.06 &       4.12 &       0.41         \\
 6731. & SII &       1.05 &       0.10 &       5.94 &       0.30         \\
 7005. & ArV & -- & -- & -- & --         \\
 7065. & HeI &      19.63 &       0.98 & -- & --         \\
 7135. & ArIII &      14.70 &       0.74 &       3.93 &       0.39         \\
 7237. & ArIV & -- & -- & -- & --         \\
 7263. & ArIV & -- & -- & -- & --         \\
 7281. & HeI &       2.18 &       0.22 & -- & --         \\
 7322. & OII &      14.60 &       0.73 &       1.70 &       0.17         \\
 7333. & OII &      13.05 &       0.65 & -- & --         \\
 7751. & ArIII & -- & -- & -- & --         \\
 9069. & SIII &     223.07 &      11.15 & -- & --         \\
 9532. & SIII &     713.24 &      35.66 & -- & --         \\
 $\log(F\mbox{H}\beta)$ &  &    -16.211 &   &    -15.441 &           \\

 \hline						     
  \end{tabular}
 \end{center}
\end{table*}

\subsubsection{Reddening correction \label{sec:reddening}}

\begin{figure} 
\includegraphics[width=8cm]{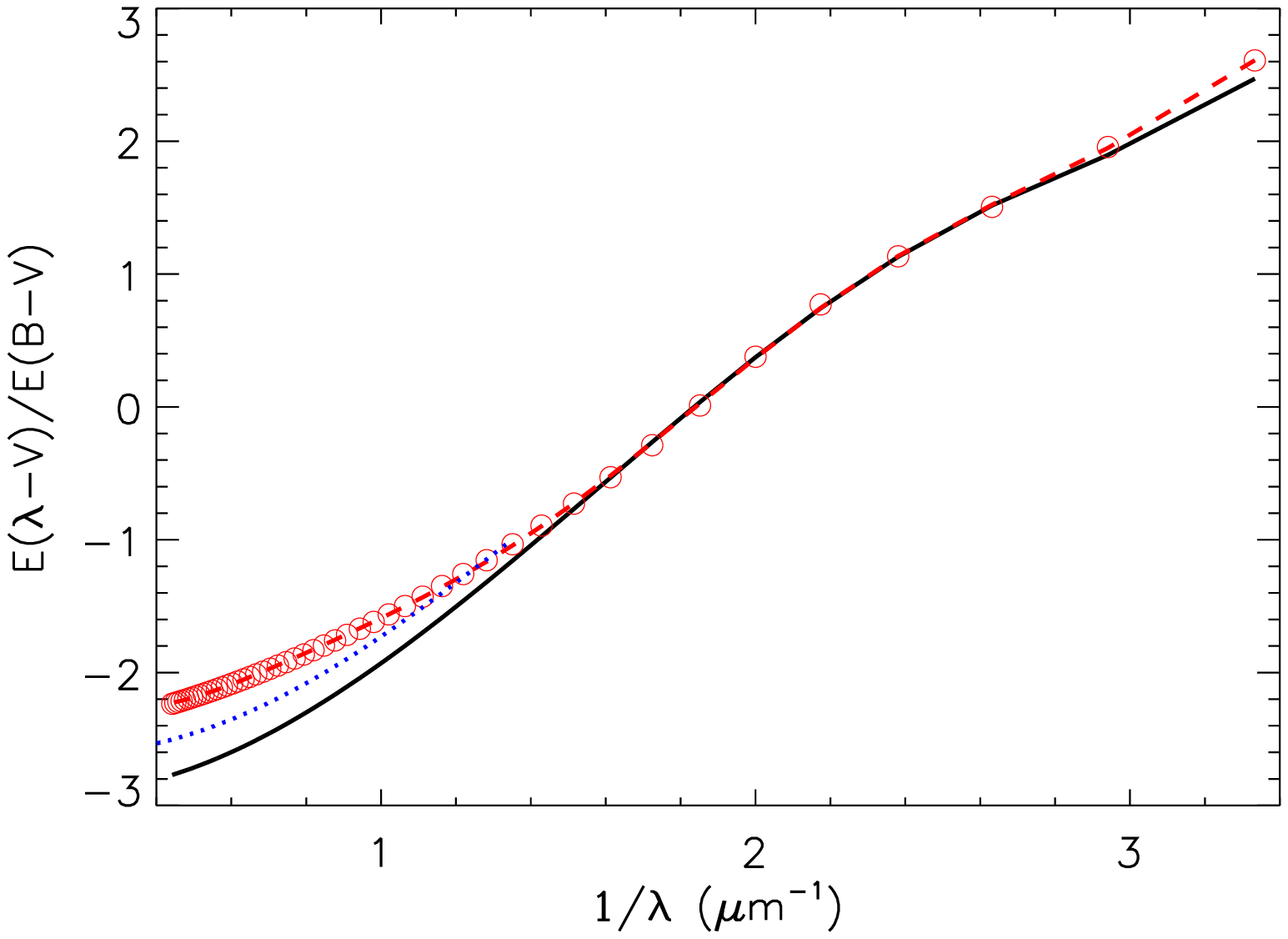}
\caption{Normalized interstellar extinction curves from NIR through optical. Open circles are the data generated using the \citet{fitzpatrick99} extinction parametrization in the case $R\equiv A(V)/E(B-V) = 2.5$. The dashed line is the seventh-degree polynomial fit. The black continuous line is the same parametrization for the case $R_V = 3.1$. The dotted line corresponds to the NIR extinction parametrization from \citet{fitzpatrick09} in the case of $R_V = 2.64$ and $\alpha = 2.49$. \label{fig:extinc}}
\end{figure}

To perform a good interstellar extinction correction is crucial for high-extinction GBPNe.  As pointed out by \citet{nataf13}, for most of the bulge, $A_V \approx 2$ is typical. However, for some regions close to the GC, $A_V \approx 50$. The extinction towards the bulge is not only high but also non-standard. The standard value of $R_V$ (the ratio of the total $A_V$ to selective $E(B-V)$ extinction at $V$) is 3.1 \citep{fitzpatrick99}. However, \citet{nataf13} find that the optical and NIR reddening law toward the inner Galaxy approximately follows an $R_V \approx 2.5$ extinction curve. On the other hand, \citet{nataf16} combined four measures of extinction in the bandpasses $V I J K_s$ and observed that there is no compatibility between bulge extinction coefficients and literature extinction coefficients using any extinction parametrization available \citep[eg.][]{cardelli89,fitzpatrick99}.
Fig. \ref{fig:extinc} shows a comparison between the extinction curves from \citet{fitzpatrick99} in the cases of $R\equiv A(V)/E(B-V) = 3.1$ and 2.5 (black continuous line and red open circles, respectively). The dashed line is the seven degree polynomial fit for $R = 2.5$. The coefficients of the parametrized extinction curve are shown in Table \ref{tab_extinc}. In the same figure, we show for comparison the NIR extinction curve parametrization from \citet{fitzpatrick09} using $R_V=2.64$ and $\alpha = 2.49$ (see their work for details). Differences between both parametrizations are noted for $\lambda > 1 \mu$m. Since the most NIR line in our data is the the [S\,{\sc iii}]$ \lambda$9532 line, we do not expect meaningful variations in the extinction-corrected lines using both parametrizations. In fact, we estimated a difference of $\sim 1$ per~cent in our approach compared with the NIR parametrization of \citet{fitzpatrick09} at 9532 \AA. Therefore, given the uncertainty in the literature with respect to the NIR ($\lambda > 7500 $ \AA) extinction, we opted to use the \citet{fitzpatrick99} extinction parametrization with $R_V=2.5$, instead of use any of the NIR parametrization provided in the literature.

Our package PNPACK is also able to perform interstellar reddening correction using both extinction curves of \citet{cardelli89} and \citet{fitzpatrick99}. As explained above, in this work we adopted the later extinction curve, since it has been show that it produces better results than the extinction curve of \citet{cardelli89} for GBPNe \citep[see e.g.][for a discussion]{escudero04}. The \citet{fitzpatrick99} extinction curve is parametrized as a function of $R_V$. Therefore, one can obtain the extinction curve computed for the case $R_V=2.5$ with a seventh-degree polynomial fitting in the form:

\begin{equation}
\frac{A_\lambda}{E(B-V)}=\sum_{n=0}^{7} a_n x^n,
\label{eq:coef_extinc}
\end{equation}
with $x=1/\lambda \ (\mu{\rm m}^{-1})$. Table \ref{tab_extinc} shows the coefficients of the polynomial fitting obtained using equation \ref{eq:coef_extinc}. 

   \begin{table}
  \begin{center}
  \caption{Coefficients of the parameterized extinction curve.}
  \label{tab_extinc}
  \setlength{\tabcolsep}{0.25em}
  \begin{tabular}{l c c c c c c c c}\hline \hline
  $n$ & 0 & 1 & 2 & 3 & 4 & 5 & 6 & 7\\ 
  $a_n$ & 1.29 & -7.80 & 20.60 & -24.84 & 16.77 & -6.21 & 1.17 & -0.09\\  
      \hline						     
  \end{tabular}
 \end{center}
\end{table}

We determined the reddening from the optical spectra, using the H$\alpha$ and H$\beta$ Balmer lines and assuming a dereddened H$\alpha$/H$\beta$ flux ratio as given by the recombination theory: I(H$\alpha$)/I(H$\beta) \sim 2.86$. Column three of Table \ref{tab_phys_param} shows the extinction $E(B-V)$ obtained for each object of our sample. The reddening showed in this table was applied both in the optical and NIR spectra. The remaining columns of Table \ref{tab_phys_param} list the electron densities and temperatures and will be introduced in Section \ref{sec:physical_parameters}.

\subsubsection{Line fluxes uncertainties}

\begin{figure}
\includegraphics[width=8cm]{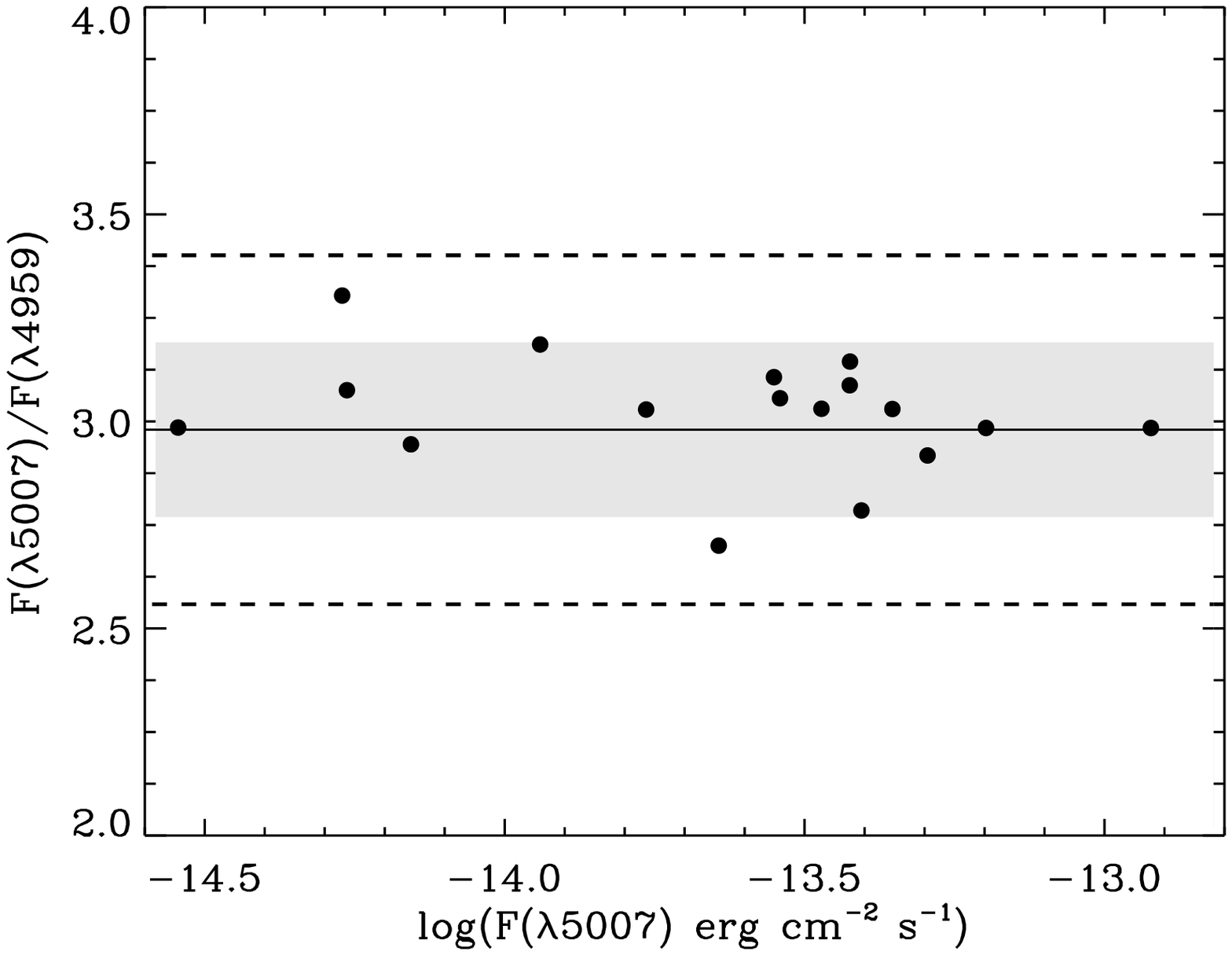}
\caption{Ratio of [O\,{\sc iii}]$\lambda$5007 and [O\,{\sc iii}]$\lambda$4959 fluxes as a function of the [O\,{\sc iii}]$ \lambda$5007 line-flux. The horizontal dashed line denotes the theoretical flux ratio [O\,{\sc iii}]$\lambda$5007/4959 of 2.98. The shaded area corresponds to the expected uncertainty in the line ratios, considering an error of 5 per cent in the line fluxes. The dashed lines corresponds to the expected uncertainty in the line ratios, considering an error of 10 per cent in the line fluxes.
\label{fig:oiii_ratio}}
\end{figure}

Sources of errors in the flux measurements can be photon shot and CCD readout noise,  bias and sky background induced noise, errors in the standard star calibration and in the interstellar and atmospheric extinction corrections. Bad sky-features subtraction or field stars contamination can also contribute to the uncertainties in the line fluxes. The quality of the data can be evaluated since some line ratios are known from atomic physics. For example, the line ratio  [O\,{\sc iii}]$\lambda$5007/4959 is expected to be 2.98 \citep{storey00}. Fig. \ref{fig:oiii_ratio} shows the [O\,{\sc iii}]$\lambda$5007/4959 ratio as a function of the [O\,{\sc iii}]$\lambda$5007 line flux. The dashed line is the theoretical value of 2.98 and the shaded area corresponds to expected uncertainty of 5 per cent around this value. The dispersion of the observed ratios in this figure is consistent with an uncertainty of 5 per cent in the flux measures. It is important to note that in cases of lines with a lower S/N or strong blending, the uncertainty is higher and can reach 40 per cent in the worst cases. In cases of strong line blending, as in the 2010 observations due to the low resolution of the 300 l/mm grating, the flux of the line [N\,{\sc ii}]$\lambda$6548 was calculated from the known ratio [N\,{\sc ii}]$\lambda$6584/6548 of 3 \citep{osterbrock06}.

\begin{figure}
\includegraphics[width=8cm]{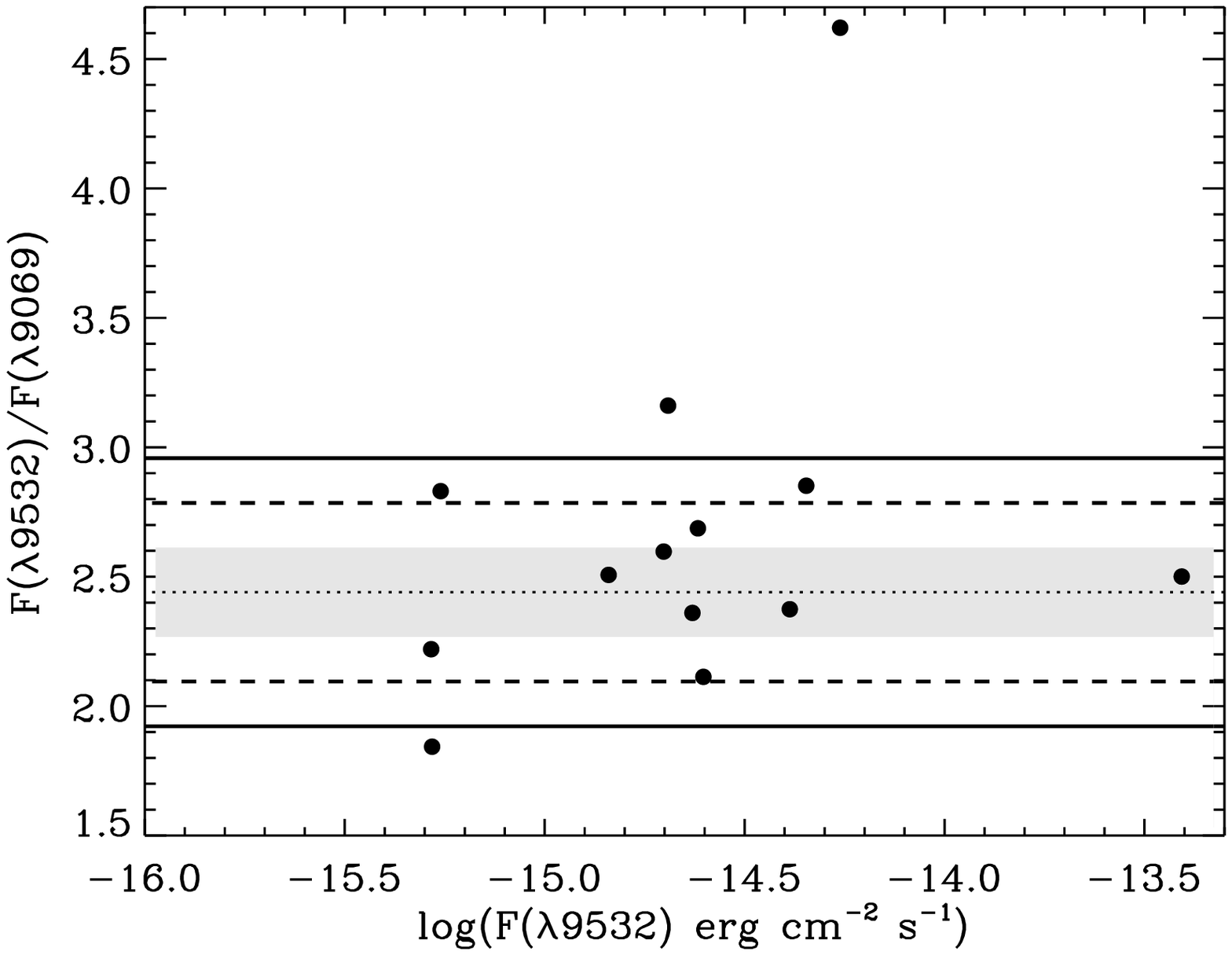}
\caption{Ratio of [S\,{\sc iii}]$\lambda$9532 and [S\,{\sc iii}]$\lambda$9069 fluxes as a function of the [S\,{\sc iii}]$ \lambda$9532 line-flux. The horizontal dotted line denotes the theoretical flux ratio of 2.44. The shaded area corresponds to the expected uncertainty in the line ratios, considering an error of 5 per cent in the line fluxes. The dashed and solid lines corresponds to the expected uncertainty in the line ratios, considering an error of 10 and 15 per cent in the line fluxes, respectively.
\label{fig6}}
\end{figure}

Fig. \ref{fig6} shows the ratio of [S\,{\sc iii}]$\lambda$9532 and [S\,{\sc iii}]$\lambda$9069 fluxes as a function of the [S\,{\sc iii}]$ \lambda$9532 line-flux. The theoretical predicted value by atomic physics of 2.44 is shown in the figure for reference. As can be seen in figures \ref{fig:oiii_ratio} and \ref{fig6}, most of the flux ratios are consistent with a line uncertainty of 5 per cent. In the case of [O\,{\sc iii}], all but two objects PPAJ1738-2800 and JaSt 98) have line ratios compatible with flux uncertainties within 10 per cent.  In the case of [S\,{\sc iii}] objects  JaSt 19, JaSt 23, JaSt 46, JaSt 52 and JaSt 77  have line ratios compatible with flux uncertainties of 5 per cent. Objects JaSt 36, JaSt 64 and JaSt 66 have flux uncertainties larger than 15 per cent. Probably a better flux calibration and spectrum extraction and/or correction for telluric lines are needed for these objects. In the case of objects having [S\,{\sc iii}] line ratios higher than 2.44, may indicate a telluric absorption in the [S\,{\sc iii}]$\lambda$9069. Those having [S\,{\sc iii}] line ratios lower then the theoretical value, may indicate an absorption in the  [S\,{\sc iii}]$ \lambda$9532. Therefore, some caution should be devoted in the electronic temperature from [S\,{\sc iii}] lines for these objects. The PN JaSt 79 shows a stellar continuum at the NIR in its spectrum that is not observed in the other PNe of our sample. Indeed, this object was classified as a symbiotic star by \citet{miszalski09}. Also, our spectrum of this object shows high ionization lines as [Fe\,{\sc Vii}] $\lambda \lambda$ 5721, 6087, with reddened fluxes $1.154\times10^{-15}$ and $3.240\times10^{-15}$ erg cm$^{-2}$ s$^{-1}$, respectively, confirming its symbiotic nature. Therefore, this object  will be excluded from the abundance analysis performed in Section \ref{sec:analysis}.

\section{Determination of physical parameters, ionic and total abundances}
\label{sec:physical_parameters}

\subsection{Physical parameters}

Table \ref{tab_phys_param} shows the electron densities derived from [S\,{\sc  ii}]$\lambda$6731/6717 line ratios. In the same table we list the electron temperatures obtained from [N\,{\sc ii}]$\lambda$5755/(6584+6548) line ratios and [S\,{\sc iii}]$\lambda$6312/(9532+9069). 

\begin{table*}
  \begin{center}
  \caption{Extinction and physical parameters.}
  \label{tab_phys_param}
  \begin{tabular}{l l c c c c c}\hline \hline
 PNG &  Name & E(B-V) &  $n_{\rm e}$ [S\,{\sc ii}]  & $T_{\rm e}$ [N\,{\sc ii}]  &  $T_{\rm e}$ [O\,{\sc iii}]  &   $T_{\rm e}$ [S\,{\sc iii}]  \\
     &       &        &   (cm$^{-3}$)       &  (K)             &      (K)           &      (K)            \\
  \hline
 
 0.344+1.567$\phantom{}^1$ & JaSt 23 &       2.94 $\pm$       0.12 &  7159  &     41566$\phantom{}^c$  &     11260  &     13122 $\pm$      1103 \\
000.2+01.7 & JaSt 19 &       2.13 $\pm$       0.12 &      1110                             &     11442 $\pm$      1095 &     11442$\phantom{}^b$ &      -- \\
000.2-01.4 & JaSt 79 &       2.31 $\pm$       0.12 &     19064 $\pm$      1299 &      20000$\phantom{}^d$ &      20000$\phantom{}^d$ &     43105$\phantom{}^c$  \\
000.4+01.1 & JaSt 36 &       2.95 $\pm$       0.15 &      3597                             &     11018 $\pm$      1552 &     11018$\phantom{}^b$ &     10296 \\
000.5+01.9 & JaSt 17 &       1.95 $\pm$       0.13 &      2077 $\pm$      1077  &     12836 $\pm$      1934 &     12836$\phantom{}^b$ &      7414  $\pm$       600 \\
000.6-01.0 & JaSt 77 &       3.46 $\pm$       0.13 &     16006                             &     11454$\phantom{}^a$  &     11454                            &     13355 $\pm$      1992 \\
000.9+01.8 & PPAJ1740-2708  &       2.56 $\pm$       0.11 &      3892 $\pm$      1111 &  9867 $\pm$  533 &      9867$\phantom{}^b$  &      -- \\
001.0+01.3 & JaSt 41 &       3.07 $\pm$       0.11 &      2336 $\pm$      1929 &     10712$\phantom{}^a$ &     10712 &     12464 $\pm$       840 \\
001.5+01.5 & JaSt 46 &       2.26 $\pm$       0.11 &      7235                             &      9653 $\pm$      1226 &     9653$\phantom{}^b$ &      7110 $\pm$       335 \\
001.7+01.3 & JaSt 52 &       2.44 $\pm$       0.12 &     11470                            &     13308 $\pm$      2154 &     13308$\phantom{}^b$  &      6948 $\pm$       592 \\
002.0-01.3 & JaSt 98 &       3.42 $\pm$       0.13 &      3035                              &     14135 $\pm$      1357 &     10258 $\pm$      1357 &     11920  \\
357.7+01.4 & PPAJ1734-3004  &       2.57 $\pm$       0.12 &      3529 $\pm$      2255 &     11665 $\pm$       970 &     11665$\phantom{}^b$  &     -- \\
358.5-01.7 & JaSt 64 &       3.06 $\pm$       0.13 &     12520                              &     15533 $\pm$      2785 &     15533$\phantom{}^b$ &     11755 $\pm$      1209 \\
358.9-01.5 & JaSt 65 &       2.98 $\pm$       0.13 &      6689                              &     20000$\phantom{}^d$ &     20000$\phantom{}^d$ &      27083$\phantom{}^c$ \\
359.5-01.2 & JaSt 66 &       3.26 $\pm$       0.12 &      2678 $\pm$      1600   &     11184$\phantom{}^a$   &     11184  &     13030 $\pm$      1361 \\
359.5-01.3 & JaSt 68 &       3.81 $\pm$       0.12 &      6701 $\pm$ -- &     13441$\phantom{}^a$  &     13441&     15739 $\pm$      2083 \\
359.9+01.8 & PPAJ1738-2800  &       1.86 $\pm$       0.13 &      2420 $\pm$       875 &     12749 $\pm$      1001 &     12749$\phantom{}^b$  &     -- \\

 \hline			
 \multicolumn{2}{l}{\footnotesize $\phantom{}^1$ OH 0.344 +1.567.}\\
 \multicolumn{2}{l}{\footnotesize $\phantom{}^a$ Adopted from $T_{\rm e}$ [O\,{\sc iii}].}\\
 \multicolumn{2}{l}{\footnotesize $\phantom{}^b$ Adopted from $T_{\rm e}$ [N\,{\sc ii}].}\\
 \multicolumn{2}{l}{\footnotesize $\phantom{}^c$ Non-physical value.}\\
 \multicolumn{2}{l}{\footnotesize $\phantom{}^d$ Upper limit adopted.}\\

  \end{tabular}
 \end{center}
\end{table*}

The [O\,{\sc iii}] temperatures were derived from a linear correlation with [S\,{\sc iii}] temperatures obtained by \citet{henry04}:

\begin{equation}
T_{e\mbox{\scriptsize [SIII]}} = -0.039(\pm 0.11)+1.20(\pm 0.11)\times T_{e\mbox{\scriptsize [OIII]}},
\label{eq:toiii}
\end{equation}
where temperatures are in units of $10^4$ K. \citet{henry04} estimated that the [S\,{\sc iii}] temperatures can be obtained from this equation with an uncertainty of $\sim 1000$ K. The $T_{\rm e\mbox{\scriptsize [OIII]}}$ temperatures calculated from this linear correlation included in table \ref{tab_phys_param} also are listed with errors. The errors were calculated from the Monte Carlo procedure explained in Section \ref{sec:errors} and only include line-flux uncertainties. We prefer to use $T_{\rm e\mbox{\scriptsize [OIII]}}$ calculated from the $T_{\rm e\mbox{\scriptsize [SIII]}}$. However, in some cases $T_{\rm e\mbox{\scriptsize [SIII]}}$ was very low to calculate $T_{\rm e\mbox{\scriptsize [OIII]}}$  from equation \ref{eq:toiii}. In these few cases, we adopted the $T_{\rm e\mbox{\scriptsize [NII]}}$ instead of $T_{\rm e\mbox{\scriptsize [OIII]}}$ and they are listed without errors in table \ref{tab_phys_param}. We derived $T_e$ from [S\,{\sc iii}] only in the cases where the line ratio 9532/9069 was within the dashed lines of Fig. \ref{fig6}. By doing this, the individual line fluxes should have errors 10 per cent or lower.

As discussed in Section \ref{sec:obs}, objects JaSt 36, JaSt41 and JaSt 64  have uncertainties in the [S\,{\sc iii}]  9532 and 9069\AA \ lines fluxes higher than 15 per cent. This can introduce higher errors in the electron temperatures obtained from [S\,{\sc iii}]. In the case of JaSt 64, we opted to use $T_{\rm e\mbox{\scriptsize [OIII]}}$ from $T_{\rm e\mbox{\scriptsize [NII]}}$.  In the case of JaSt 41, JaSt 66 where the observed ratio was lower than the theoretical one, indicating a stronger absorption in the 9532 line, the total dereddened [S\,{\sc iii}] intensity  was taken equal to 2.44 times the 9069 line. In the remaining cases (JaSt 17, JaSt 36 and JaSt 98) where the ratio was larger than the theoretical one, the 9532 line was taken as reference, instead. Examining in detail the spectra of JaSt 41, we noted a contamination by a field star. Therefore some caution should be taken with the results for this object. 

In the case of object JaSt 65, we obtained a non-physical value of  27083 K for the temperature from [S\,{\sc iii}] lines. This temperature is not typical for PNe and to calculate the abundances for this object we adopted an upper limit of $2\times10^4$ K for  $T_{\rm e\mbox{\scriptsize [OIII]}}$ and $T_{\rm e\mbox{\scriptsize [NII]}}$. However, since the abundances derived from collisional excitation lines depend strongly on the temperatures, the uncertainties in the abundances for this object are very high. Therefore, this object will not be used in the abundance analysis. JaSt 79 was also excluded from the abundance analysis because of its symbiotic nature. This can be noted in table \ref{tab_phys_param}, where the derived [S\,{\sc iii}] electron temperature yielded 43105 K, a very high value for a typical PNe. The electronic temperature from  [N\,{\sc ii}] for JaSt 23 is also non-typical for PNe. Nonetheless, $T_{\rm e\mbox{\scriptsize [SIII]}}$ resulted in a plausible value. In view of the discrepancy between both values, some caution should be taken with the chemical abundances of this object and we prefer not use its abundances in the abundance analysis.

\subsection{Ionic abundances}

\begin{figure}
\includegraphics[width=8cm]{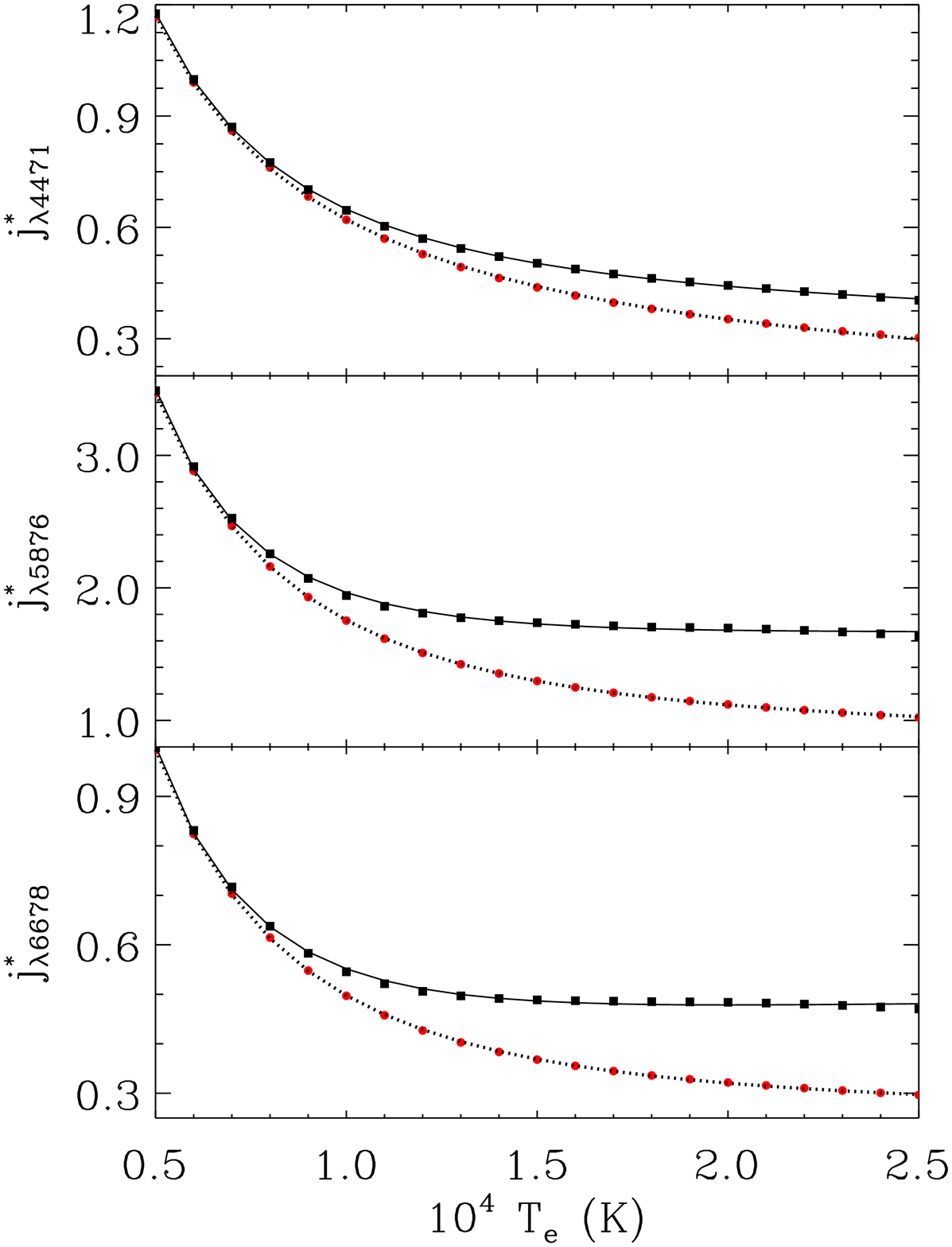}
\caption{Emissivities of He\,{\sc i} from \citet{porter13} as a function of the electron temperature $T_{\rm e}$ for two different values of electron density and fitted functions:  $10^3$ cm$^{-3}$ (red circles, dotted lines) and $10^4$ cm$^{-3}$ (black squares, continuous lines). The emissivities are expressed as $j^*_\lambda = 4\upi j_\lambda /n_{\rm e}n_{{\rm He}^+}$ in units of $10^{-25}$ erg cm$^3$ s$^{-1}$ . \label{fig7}}
\end{figure}

The abundance for He$^+$ was derived from the average of the recombination lines 4471, 5876 and 6678 \AA \ . The average was weighted by the intensity of each line. New He\,{\sc i} emissivities have recently become available through the work of \citet{porter12} and recently corrected by \citet{porter13}. These are the most recent He\,{\sc i} emissivities, and collisional effects are already included in the emissivities calculation. In this work the emissivities of \citet{porter13} are adopted in order to calculate the He\,{\sc i} abundances. Their emissivities are tabulated for discrete of electron densities and temperatures. Therefore, we fitted the values provided by \citet{porter13} for $n_{\rm e} = 10^3 \mbox{ and } 10^4$ cm$^{-3}$ and $T_{\rm e}$ between $5\times10^3$ and $2.5\times10^{4}$ K, using the following parametrization:

\begin{equation}
 \frac{4\pi j_\lambda}{n_{\rm e} n_{\mbox{\scriptsize He}^+}} = [a + b(\mbox{ln } T_{\rm e})^2 + c \mbox{ ln }T_{\rm e} + \frac{d}{\mbox{ln }T_{\rm e}}] T_{\rm e}^{-1},
 \label{eq:heI_em} 
\end{equation}
in units of $10^{-25} \ {\rm ergs } \ {\rm cm }^3 \ {\rm s}^{-1}$. The results are shown in Fig. \ref{fig7}, where the He\,{\sc i} emissivities for the lines 4471, 5876 and 6678 \AA \ are displayed as a function of the electron temperature, for two values of electron densities: $10^3$ cm$^{-3}$ (red circles) and $10^4$ cm$^{-3}$ (black squares). Fits using equation \ref{eq:heI_em} are displayed at the same figure as dotted lines ($n_{\rm e} = 10^3$ cm$^{-3}$) and continuous lines ($n_{\rm e} = 10^4$ cm$^{-3}$). The coefficients of the fits using equation \ref{eq:heI_em} are shown in Table \ref{tab:heI_em}.

\begin{table}
  \small
  \begin{center}
  \caption{Coefficients for the  He\,{\sc i} emissivities.}
  \label{tab:heI_em}
  \begin{tabular}{l c c c c}\hline \hline
  $\lambda$ (\AA) & $a \times 10^6$ & $b\times10^3$ & $c \times 10^5$  & $d \times 10^6$\\
  \hline
   &  & $n_{\rm e} = 10^3$ cm$^{-3}$ &  &  \\
  \hline 
  4471 & 0.52235 & -2.39494 & -0.60726 & -1.47358 \\
  5876 & 7.13671 & 31.32693 & -8.18163 & -20.64105 \\
  6678 & 2.17062  & 9.55600 & -2.49280 & -6.26613 \\   
   \hline						     
     &  & $n_{\rm e} = 10^4$ cm$^{-3}$ &  &  \\
  \hline 
 4471 & 2.00719 & 9.16167 & -2.34319 & -5.70799 \\
  5876 & 9.73706 & 46.87604 & -11.70238 & -26.85340\\
  6678 & 3.31042  & 15.59403 & -3.93518 & -9.24089 \\   
   \hline						     
  \end{tabular}
 \end{center}
\end{table}

Values between the fitted functions displayed in Fig. \ref{fig7} were interpolated on a logarithmic scale following \citet{monreal-ibero13} as:

\begin{equation}
 j^*_\lambda[\log(n_{\rm e})]=[ j^*_\lambda(4) -  j^*_\lambda(3)]\times [\log(n_{\rm e}) - 3]+  j^*_\lambda(3),
\end{equation}
where we defined $ j^*_\lambda = 4\upi j_\lambda/n_{\rm e} n_{\mbox{\scriptsize ion}}$, with $n_{ion} = n_{\mbox{\scriptsize He}^+}$ in this particular case.

In our code, the He$^{+2}$ abundance is calculated from the He\,{\sc ii}$\lambda4686$ \AA \ recombination line. The recombination coefficients tabulated in \citet{osterbrock06} were used as well as the H$\beta$ emissivity provided by \citet{aver10}. We fitted the ratio  $j^*_{\lambda4861}/j^*_{\lambda4686}$ with a second-degree polynomial function. For typical densities found in PNe, this ratio do not depend strongly on the density, so that we adopted $n_{\rm e} = 10^4 \mbox{ cm}^{-3}$ and the result of the fit is shown in equation \ref{eq:heii_fit}.  Both He$^{+}$ and He$^{+2}$ abundances are calculated using the O\,{\sc iii} electron temperature.

\begin{equation}
 \frac{j^*_{\lambda4861}}{j^*_{\lambda4686}}=5.1340 - 5.062\times10^{-4} \ T_{\rm e} + 1.4280\times10^{-8} \ T_{\rm e}^2
 \label{eq:heii_fit}
\end{equation}

For collisionally excited lines, we calculated the ionic abundances using the {\it nebular} software \citep{shaw95} and the appropriate electron temperature for the corresponding ionization zone. 

\begin{table*}
  \begin{center}
  \caption{Ionic abundances.}
  \label{tab_ionic_abund}
  \begin{tabular}{l c c c c c c c c}\hline \hline
 PNG & He$^+$ & He$^{++}$ & O$^+$ & O$^{++}$ & N$^+$ & S$^+$ & S$^{++}$ & Ar$^{++}$\\
 
         &        &      & $\times 10^{-4}$ & $\times 10^{-6}$ & $\times 10^{-6}$ & $\times 10^{-6}$ & $\times 10^{-6}$ & $\times 10^{-6}$ \\ 
  \hline
  
0.344+1.567$\phantom{}^1$ &        0.058 $\pm$      0.007 & --   &       0.66   &       0.06   &      33.57   &       0.27   &       7.30   &       0.66   \\
000.2+01.7 &        0.075 $\pm$      0.010 &      0.018 $\pm$      0.003 &       0.08   &       2.28   &       0.44 $\pm$       0.15 &       0.04 $\pm$       0.01 &       0.55   &       0.64   \\
000.2-01.4 &        0.034 $\pm$      0.004 &      0.059 $\pm$      0.011 &       0.08 $\pm$       0.01 &       0.45 $\pm$       0.05 &       2.75 $\pm$       0.30 &       0.13 $\pm$       0.02 &       1.74   &       0.70 $\pm$       0.06 \\
000.4+01.1 &        0.117 $\pm$      0.016 & --   &       0.14   &       3.11 $\pm$       1.39 &       2.90 $\pm$       1.18 &       0.11 $\pm$       0.06 &       2.35 $\pm$       1.50 &       0.57 $\pm$       0.21 \\
000.5+01.9 &        0.143 $\pm$      0.018 & --   &       0.05   &       0.79 $\pm$       0.39 &       3.09 $\pm$       1.23 &       0.06 $\pm$       0.02 &       0.53   &       0.73 $\pm$       0.23 \\
000.6-01.0 &       0.121 $\pm$      0.013 & --   &       0.33   &       2.94   &       2.01   &       0.17   &       2.36   &       0.84   \\
000.9+01.8 &        0.165 $\pm$      0.017 &      0.019 $\pm$      0.003 &       0.94 $\pm$       0.34 &       3.60 $\pm$       0.78 &     126.05 $\pm$      23.66 &       2.88 $\pm$       0.76 &       7.75 $\pm$       2.39 &       3.56 $\pm$       0.65 \\
001.0+01.3 &        0.131 $\pm$      0.016 & --   &       0.32   &       1.25   &       3.45   &       0.06   &       2.03   &       0.32   \\
001.5+01.5 &        0.134 $\pm$      0.017 & --   &       0.29 $\pm$       0.24 &       1.85 $\pm$       0.89 &      31.55 $\pm$      14.32 &       0.61   &       3.91   &       2.23 $\pm$       0.70 \\
001.7+01.3 &        0.049 $\pm$      0.006 & --   &       0.53 $\pm$       0.37 &       0.14 $\pm$       0.06 &       9.93 $\pm$       5.30 &       0.21   &       0.50   &       0.34 $\pm$       0.11 \\
002.0-01.3 &        0.130 $\pm$      0.018 & --   &       0.17 $\pm$       0.07 &       2.15 $\pm$       0.52 &       7.17 $\pm$       1.77 &       0.06 $\pm$       0.03 &       0.93   &       0.97 $\pm$       0.21 \\
357.7+01.4 &        0.113 $\pm$      0.015 & --   &       0.21 $\pm$       0.09 &       2.19 $\pm$       0.59 &      22.76 $\pm$       5.77 &       0.79   &       6.14 $\pm$       1.95 &       0.96 $\pm$       0.23 \\
358.5-01.7 &        0.083   & --   &       0.06   &       0.76   &       3.16   &       0.12   &       1.31   &       0.51   \\
358.9-01.5 &        0.076 $\pm$      0.011 & --   &       0.10   &       0.26   &       0.67   &       0.01   &       0.71   &       0.22   \\
359.5-01.2 &        0.120 $\pm$      0.013 & --   &       0.11   &       1.03   &       2.97   &       0.06   &       1.32   &       0.22   \\
359.5-01.3 &        0.115   & --   &       0.08   &       1.15   &       1.18   &       0.05   &       2.20   &       0.46   \\
359.9+01.8 &        0.123 $\pm$      0.013 & --   &       0.25 $\pm$       0.09 &       1.04 $\pm$       0.22 &      47.11 $\pm$      10.15 &       2.14 $\pm$       0.49 &       2.52 $\pm$       0.79 &       1.79 $\pm$       0.40 \\

 \hline						     
  \end{tabular}
 \end{center}
\end{table*}

\subsection{Elemental abundances \label{subsec:elem_abund}} 

Since the spectral range of the observations is not sufficient to observe all the necessary lines of a given ion, and it is not possible to calculate the total abundance of a particular element by the direct sum of the ionic abundances of all the ions present in a nebula. Instead, it must be calculated by means of the ionization correction factors (ICFs). One of most frequently used ICFs in the literature are those from \citet{kingsburgh94}. However, recently \citet{delgado-inglada14} have published new ICFs formulae based on a grid of photoionization models and they have computed analytical expressions for the ICFs of He, O, N, Ne, S, Ar, Cl and C. According to their work, the oxygen abundances are not expected to be very different from those calculated with the ICFs of \citet{kingsburgh94}. On the other hand, the abundances of N, S, Ar, Ne calculated with the new ICFs show significant differences. A direct comparison between the abundances calculated with both ICFs is beyond the scope of this paper and the reader is referred to the original paper of \citet{delgado-inglada14}, where some comparisons are done between both ICFs. However, one should mention that a direct comparison between the abundances of N, S, Ar and Ne calculated with both ICFs is missing in that paper.

In our code PNPACK we have implemented the calculation of the elemental abundances by using the ICFs provided by \citet{delgado-inglada14} and also \citet{kingsburgh94}. In this paper, we are adopting the new ICFs proposed by \citet{delgado-inglada14}, since they incorporate more recent physics and  are derived using a wider range of parameters of photoionization models than the previous ICFs from \citet{kingsburgh94}. One advantage of the new ICFs provided by \citet{delgado-inglada14} is the possibility to compute the errors in the elemental abundances introduced by the adopted ICF approximation. They provide analytical formulae to estimate error bars associated with the ICFs, something not possible until their work. In the case of helium we opted to do not use any ICF, since the relative populations of helium ions depend essentially on the effective temperature of the central star. So that, there is no reliable way to correct for neutral helium in our objects.

The reader should to note that in some cases, where the ionic abundance was available, the elemental abundance of the corresponding ion was not possible to calculate, since the ICFs of \citet{delgado-inglada14} are not valid for the specific $v$ and $w$ parameters \citep[see][for more details]{delgado-inglada14}. In these few cases, in order to obtain the elemental abundances one needs to perform detailed photoionization models of the object. The obtained ionic abundances relative to hydrogen and uncertainties are shown in table \ref{tab_ionic_abund}.

\begin{figure}
\includegraphics[width=8cm]{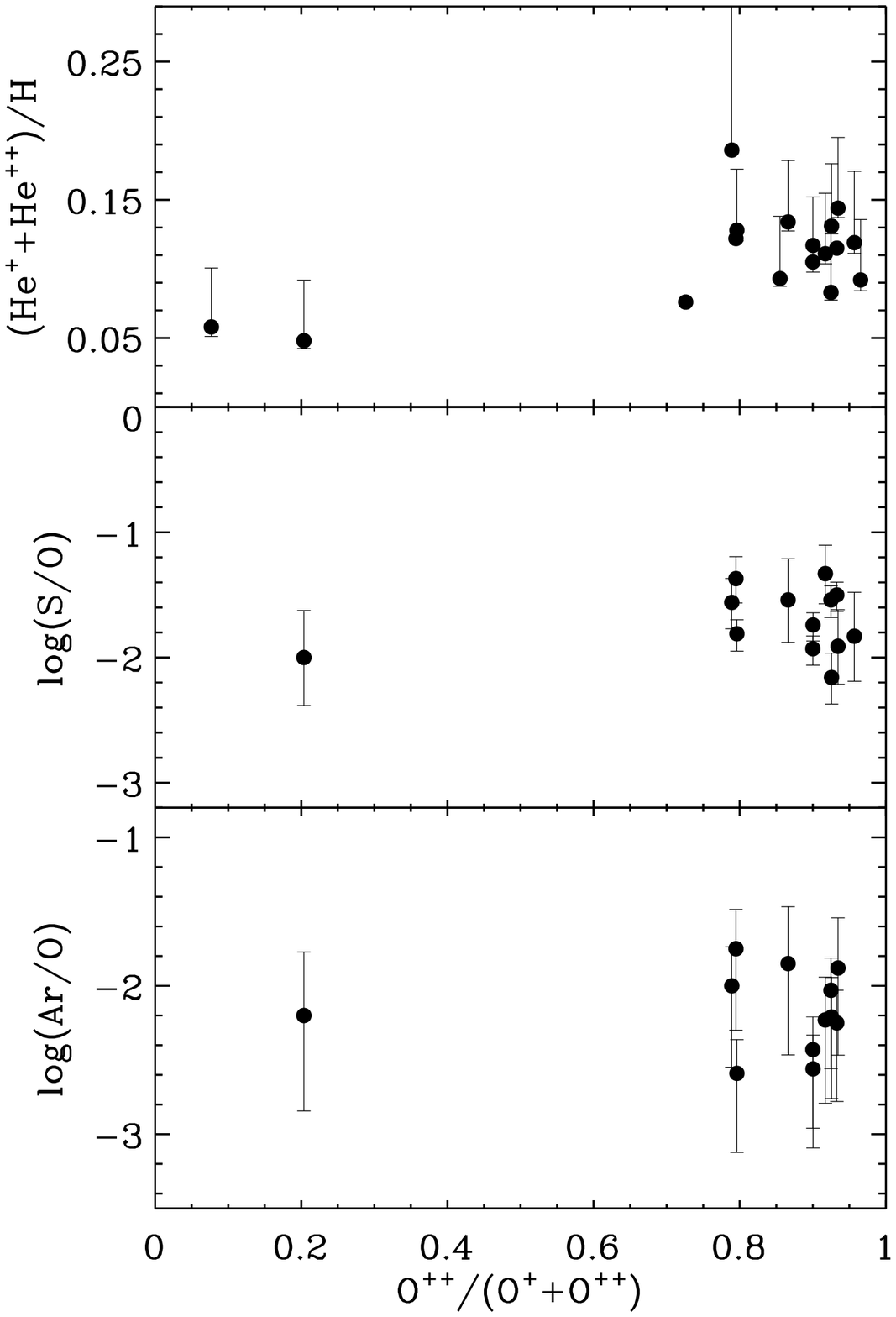}
\caption{Abundance ratios of $(\mbox{He}^+ + \mbox{He}^{++})/$H, S/O and Ar/O as a function of $\mbox{O}^{++}/(\mbox{O}^+ + \mbox{O}^{++})$ in top, middle and bottom panels, respectively.  \label{fig8}}
\end{figure}

In order to test the ICFs, Fig. \ref{fig8} shows the values of $(\mbox{He}^+ + \mbox{He}^{++})/$H, S/O and Ar/O as a function of $\mbox{O}^{++}/(\mbox{O}^+ + \mbox{O}^{++})$. No significant trend is seen for S/O and Ar/O. For $(\mbox{He}^+ + \mbox{He}^{++})/$H, since for many PNe we have not detected the presence of $\mbox{He}^{++}$ lines, the helium abundances are a lower limit for theses objects. For low excitation objects the uncertainties in the helium abundances are very high, as showed by the error bars in the figure. This can be attributed to neutral helium, as it is not taken into account by the helium ICF. 

The obtained chemical abundances and uncertainties (see next section) of He, O, N, S, and Ar with respect to H are listed in table \ref{tab_elem_abund}. 

\begin{table*}
  \begin{center}
  \caption{Elemental abundances.}
  \label{tab_elem_abund}
  \def\arraystretch{1.5}
  \begin{tabular}{l l c c c c c}\hline \hline
 PNG  & Name &  He/H & 12+$\log$ O/H & 12+$\log$ N/H & 12+$\log$ S/H & 12+$\log$ Ar/H \\
\hline

0.344+1.567$\phantom{}^1$ & JaSt 23 &      0.058   $^{+     0.326}_{-     0.003}$ &        7.85   $^{+      0.16}_{-      0.16}$ &        7.61   $^{+      0.17}_{-      0.17}$ &        6.87   $^{+      0.18}_{-      0.20}$ &        6.11   $^{+      0.26}_{-      0.55}$ \\
000.2+01.7 & JaSt 19 &      0.092   $^{+     0.042}_{-     0.005}$ &        8.43   $^{+      0.13}_{-      0.13}$ &  --     &        6.13   $^{+      0.17}_{-      0.19}$ &  --     \\
000.2-01.4 & JaSt 79 &      0.093   $^{+     0.047}_{-     0.006}$ &        8.04   $^{+      0.12}_{-      0.15}$ &        7.49   $^{+      0.22}_{-      0.22}$ &        6.71   $^{+      0.15}_{-      0.19}$ &        6.29   $^{+      0.25}_{-      0.55}$ \\
000.4+01.1 & JaSt 36 &      0.119   $^{+     0.043}_{-     0.007}$ &        8.52   $^{+      0.20}_{-      0.20}$ &        8.44   $^{+      0.31}_{-      0.30}$ &        6.69   $^{+      0.29}_{-      0.30}$ &  --     \\
000.5+01.9 & JaSt 17 &      0.144   $^{+     0.044}_{-     0.008}$ &        7.92   $^{+      0.19}_{-      0.19}$ &        8.27   $^{+      0.28}_{-      0.27}$ &        6.01   $^{+      0.22}_{-      0.24}$ &        6.04   $^{+      0.28}_{-      0.56}$ \\
000.6-01.0 & JaSt 77 &      0.105   $^{+     0.045}_{-     0.006}$ &        8.51   $^{+      0.05}_{-      0.05}$ &        7.88   $^{+      0.24}_{-      0.23}$ &        6.58   $^{+      0.09}_{-      0.12}$ &        6.08   $^{+      0.21}_{-      0.53}$ \\
000.9+01.8 & PPAJ1740-2708 &      0.186   $^{+     0.052}_{-     0.008}$ &        8.68   $^{+      0.11}_{-      0.11}$ &        8.80   $^{+      0.20}_{-      0.19}$ &        7.12   $^{+      0.16}_{-      0.18}$ &        6.68   $^{+      0.24}_{-      0.54}$ \\
001.0+01.3 & JaSt 41 &      0.128   $^{+     0.051}_{-     0.007}$ &        8.20   $^{+      0.06}_{-      0.06}$ &        7.74   $^{+      0.20}_{-      0.19}$ &        6.39   $^{+      0.09}_{-      0.13}$ &        5.61   $^{+      0.22}_{-      0.53}$ \\
001.5+01.5 & JaSt 46 &      0.134   $^{+     0.047}_{-     0.007}$ &        8.33   $^{+      0.21}_{-      0.21}$ &        8.94   $^{+      0.32}_{-      0.31}$ &        6.79   $^{+      0.25}_{-      0.27}$ &        6.48   $^{+      0.32}_{-      0.58}$ \\
001.7+01.3 & JaSt 52 &      0.048   $^{+     0.173}_{-     0.003}$ &        7.85   $^{+      0.24}_{-      0.24}$ &        7.26   $^{+      0.31}_{-      0.31}$ &        5.85   $^{+      0.29}_{-      0.30}$ &        5.65   $^{+      0.35}_{-      0.60}$ \\
002.0-01.3 & JaSt 98 &      0.131   $^{+     0.044}_{-     0.008}$ &        8.38   $^{+      0.12}_{-      0.12}$ &        8.59   $^{+      0.26}_{-      0.25}$ &        6.22   $^{+      0.15}_{-      0.17}$ &        6.17   $^{+      0.24}_{-      0.54}$ \\
357.7+01.4 & PPAJ1734-3004 &      0.111   $^{+     0.044}_{-     0.006}$ &        8.38   $^{+      0.12}_{-      0.12}$ &        9.02   $^{+      0.25}_{-      0.24}$ &        7.05   $^{+      0.19}_{-      0.21}$ &        6.15   $^{+      0.26}_{-      0.55}$ \\
358.5-01.7 & JaSt 64 &      0.083   $^{+     0.044}_{-     0.006}$ &        7.92   $^{+      0.05}_{-      0.05}$ &        8.22   $^{+      0.22}_{-      0.21}$ &        6.38   $^{+      0.10}_{-      0.13}$ &        5.89   $^{+      0.21}_{-      0.53}$ \\
358.9-01.5 & JaSt 65 &      0.076   $^{+     0.056}_{-     0.005}$ &        7.56   $^{+      0.01}_{-      0.01}$ &        6.85   $^{+      0.17}_{-      0.15}$ &        5.88   $^{+      0.05}_{-      0.10}$ &        5.40   $^{+      0.33}_{-      0.58}$ \\
359.5-01.2 & JaSt 66 &      0.117   $^{+     0.045}_{-     0.006}$ &        8.06   $^{+      0.05}_{-      0.05}$ &        8.05   $^{+      0.23}_{-      0.22}$ &        6.32   $^{+      0.08}_{-      0.12}$ &        5.50   $^{+      0.22}_{-      0.53}$ \\
359.5-01.3 & JaSt 68 &      0.115   $^{+     0.044}_{-     0.007}$ &        8.09   $^{+      0.05}_{-      0.05}$ &        7.84   $^{+      0.27}_{-      0.26}$ &        6.59   $^{+      0.09}_{-      0.12}$ &        5.84   $^{+      0.21}_{-      0.53}$ \\
359.9+01.8 & PPAJ1738-2800 &      0.122   $^{+     0.051}_{-     0.006}$ &        8.10   $^{+      0.10}_{-      0.10}$ &        8.86   $^{+      0.21}_{-      0.20}$ &        6.73   $^{+      0.14}_{-      0.17}$ &        6.35   $^{+      0.24}_{-      0.54}$ \\

 \hline						     
  \end{tabular}
 \end{center}
\end{table*}

\subsection{Uncertainties \label{sec:errors}}

Uncertainties in the abundances are due to uncertainties in the line fluxes of the abundance diagnostic lines, uncertainties in the diagnostic lines for electron temperature and density, and uncertainties in the ICF that we have assumed for the determination of elemental abundances. 

We have computed errors in the abundances and physical parameters, as well as extinction, performing a Monte Carlo procedure. We assumed for each line flux a Gaussian distribution centred at the flux effectively measured and having a dispersion equal to the estimated flux uncertainty.  The latter was calculated by comparing the line intensity with that from the H$\beta$ line. By inspection of our spectra and figures \ref{fig:oiii_ratio} and \ref{fig6}, we have adopted errors of 5 per cent for lines where the relative flux of the line with H$\beta$ ($F(\lambda)/F(\mbox{H}\beta)$) was higher than 500. Errors of 10 per cent for those ratios higher than 10 and lower than 500. An uncertainty of 20 per cent was attributed for ratios $F(\lambda)/F(\mbox{H}\beta)$ higher than 2 and lower than 10. And finally, errors of 40 per cent for lines where $F(\lambda)/F(\mbox{H}\beta)$ was lower than 2. 

\begin{figure}
\includegraphics[width=8cm]{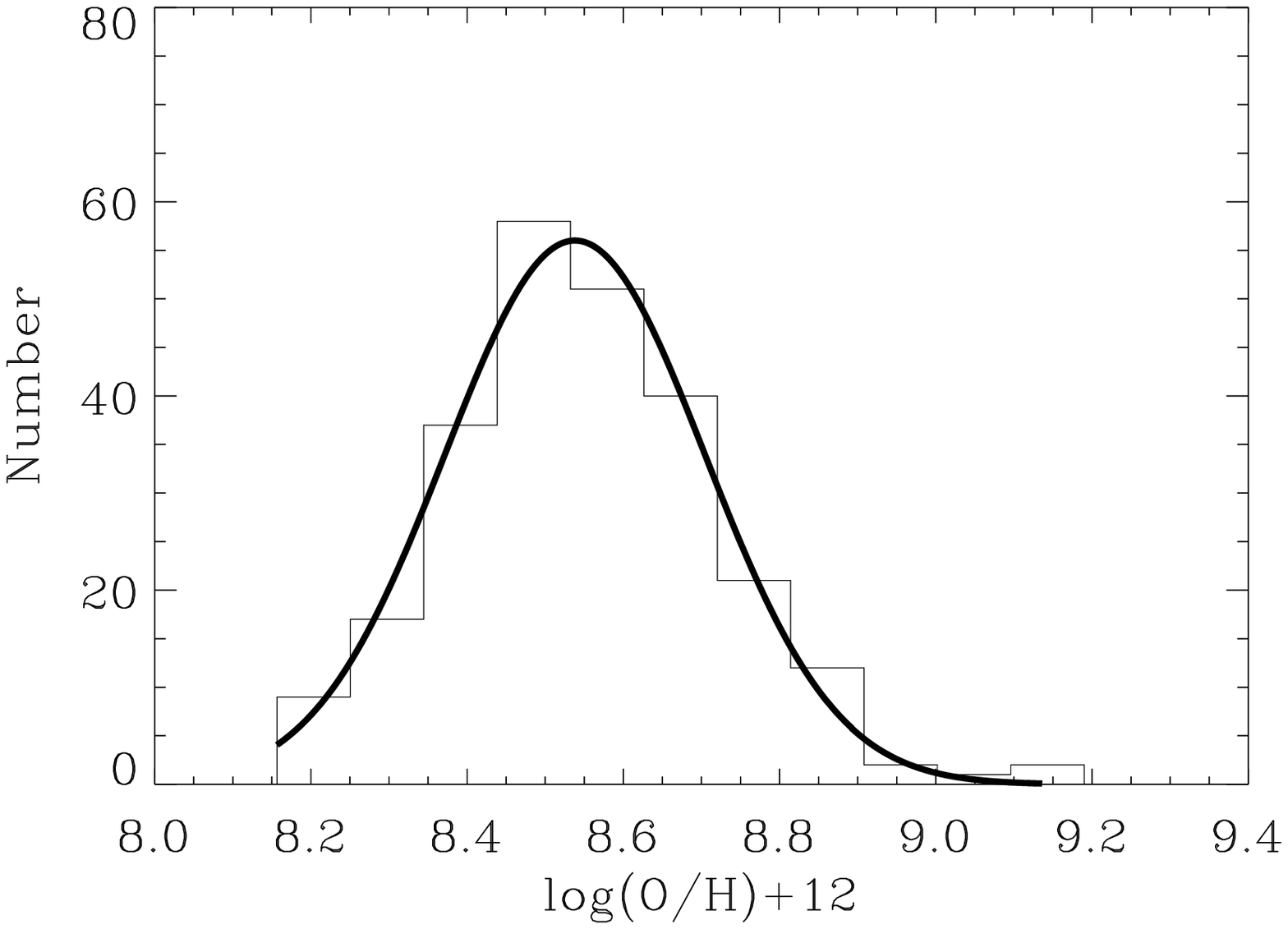}
\caption{Histogram of $\log(\mbox{O}/\mbox{H})+12$ values of the 250 independent realizations from the Monte Carlo simulation for JaSt~36. The continuous curve is the Gaussian fitted to the histogram data.\label{fig9}}
\end{figure}

For each line, we considered 250 independent realizations and fitted the histogram of the distribution with an Gaussian function. The error bars and the most probable values were estimated from the Gaussian standard error and mean, respectively, obtained from the Gaussian fit. In the cases where the distribution was clearly no Gaussian, as in most cases of electron densities, we were not able to obtain the errors. These cases are marked with an $-$ signal in tables \ref{tab_phys_param}, \ref{tab_ionic_abund} and \ref{tab_elem_abund}. An example of a Gaussian fit to the histogram of the Monte Carlo simulation for PN JaSt 36 can be seen in Fig. \ref{fig9}.

The errors in the ICFs were computed through the recent work from \citet{delgado-inglada14}, where the authors have evaluate the uncertainties in the ICFs. Thanks to their work, it is now possible to include these uncertainties in the elemental abundances computed with these ICFs. The final uncertainties on the elemental abundances were calculated by adding in quadrature the uncertainties obtained from the ICFs with the uncertainties obtained from the Monte Carlo simulations. Note that, since the uncertainties in the ICFs are not symmetrical, error bars are also not symmetrical for the final abundances.

\section{Abundance analysis}
\label{sec:analysis}

In order to seek for differences in the chemical enrichment of GBPNe near the GC and other regions of the Galactic bulge, we compared in this section the abundances obtained in the present work with the GBPNe abundances from our previous work (CCM10), where we performed spectrophotometric observations of GBPNe located in the outer regions of the bulge. Hence, a comparison between both samples can give information about the chemical enrichment of the central parts of the Milky Way galaxy. We should note that the elemental abundances from CCM10 were calculated with the ICFs from \citet{kingsburgh94} and ours with \citet{delgado-inglada14}. To provide a meaningful comparison, we recalculated the abundances from CCM10 with the new ICFs from \citet{delgado-inglada14} and also the He\,{\sc i} emissivities from \citet{porter13}. In this way, we guarantee the same methodology for both samples. 

\begin{figure}
\includegraphics[width=8cm]{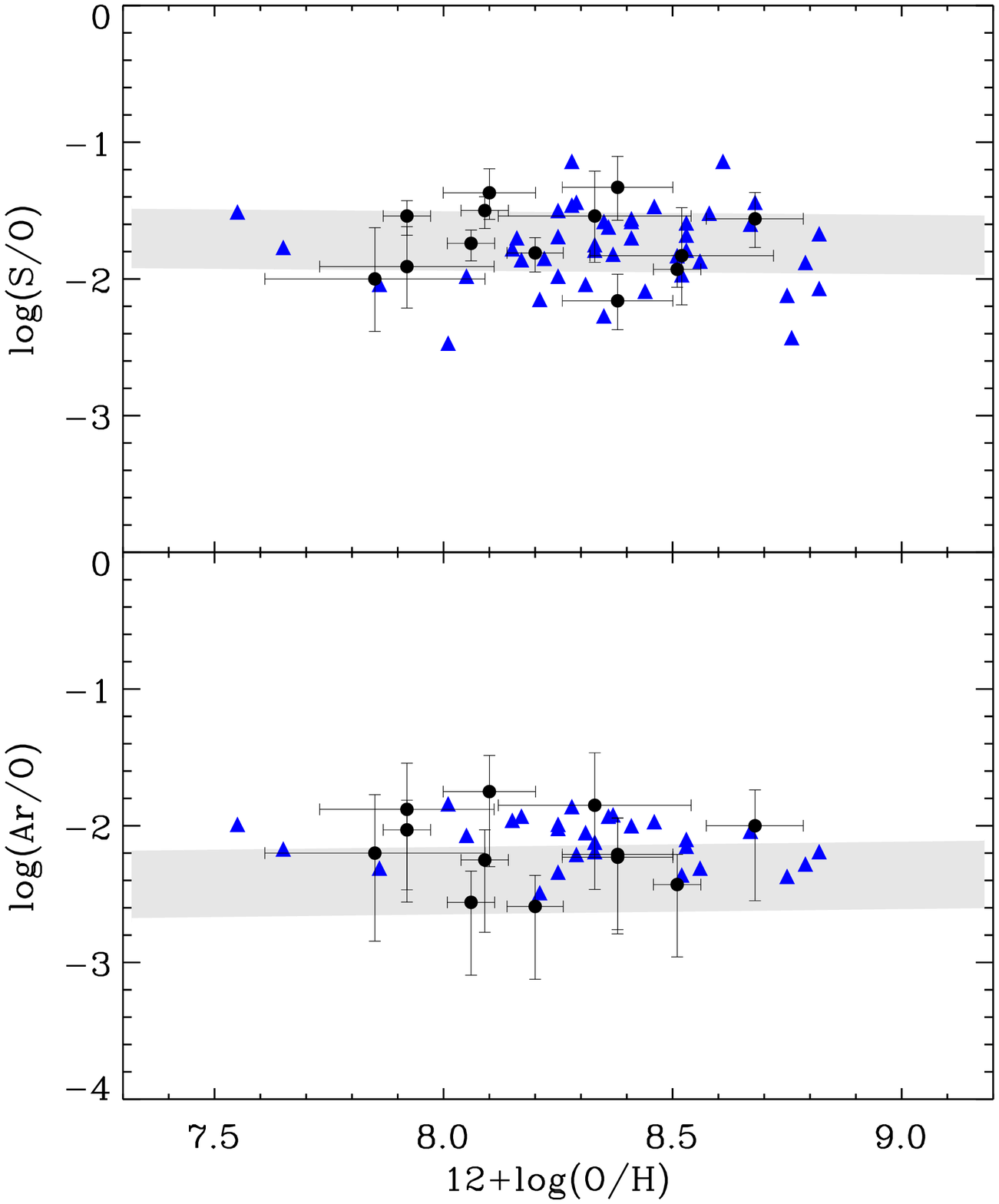}
\caption{Abundances ratios of S/O and Ar/O as a function of O/H abundances. Black filled circles with error bars are GBPNe data from this work. The GBPNe data from CCM10 are represented by blue filled triangles. The light-grey bands for $\alpha$-elements to oxygen ratio represent the 1-$\sigma$ dispersion of the relations from \citet{izotov06} for low-metallicity blue compact dwarf galaxies. \label{fig10}}
\end{figure}

Fig. \ref{fig10} shows the abundances ratios S/O and Ar/O as a function of O/H abundances for our data and our previous work (CCM10). The 1-$\sigma$ dispersion of the relations from \citet{izotov06} for low-metallicity blue compact dwarf galaxies is displayed in the same figure. The GBPNe data seems to follow the \citet{izotov06} relations, with most of them inside the 1-$\sigma$ dispersion. We note that our data for the Ar/O ratio show a higher dispersion  than the S/O ratio. Our Ar abundances are calculated in most cases from the [Ar\,{\sc iii}]  7005 and 7751\AA \ lines. Since the line 7751\AA \ is located at the IR region of the spectra where the Goodman's CCD has more fringes, this can contribute to the higher errors found for Ar abundances. Another source of error is the ICF of Ar, whose uncertainty is much higher than the other elements, as noted by \citet{delgado-inglada14}. The S/O ratio is much more in agreement with the blue compact dwarf galaxies data. It is interesting to note that in our data we do not find the sulfur anomaly problem reported by other studies in the literature \citep[see e.g.][]{henry12}. Since we have measured for most PNe of our sample the near IR [S\,{\sc iii}] $\lambda$9069 and 9532\AA \ lines, this reduce the uncertainties in S abundances, improving the accuracy of the abundances.  

\begin{figure}
\includegraphics[width=8.5cm]{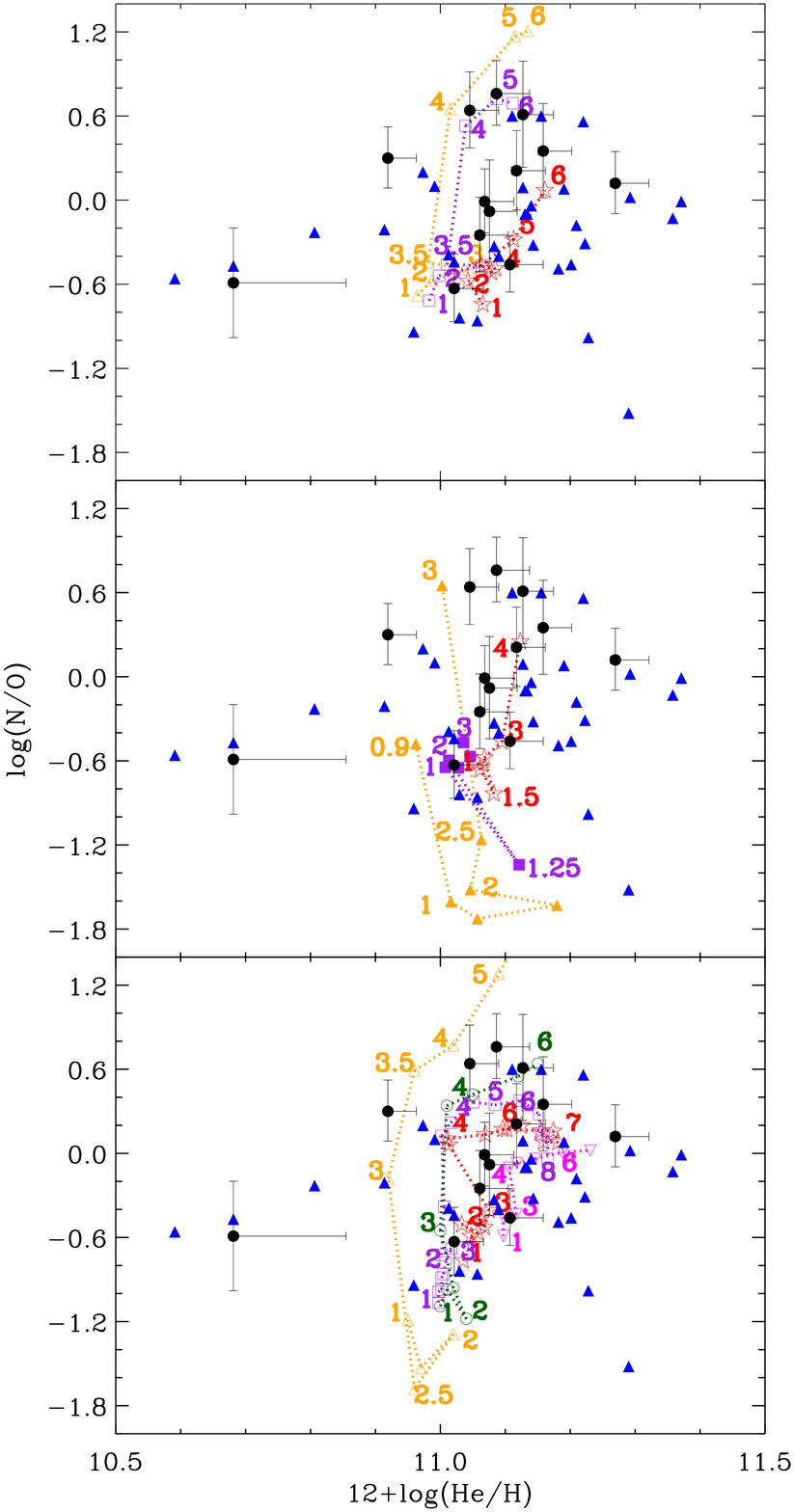}
\caption{Abundance ratio of N/O as a function of $12+\log(\mbox{He/H})$ abundances. Filled circles with error bars are data from present work, while filled blue triangles are the data from CCM10. The unfilled symbols with numbers joined by dotted lines represent the results of the AGB nucleosynthesis models and the numbers give the initial masses of the individual models in M$_{\sun}$ units. Top: Models from \citet{karakas10} for a given value of $Z$ as orange triangles for $Z=0.004$, purple squares for $Z=0.008$ and red stars for $Z=0.02$; Middle: models from \citet{miller16} as orange triangles for $Z=0.001$, purple squares for $Z=0.01$ and red circles for $Z=0.02$; Bottom: models from \citeauthor{ventura14a} group as orange triangles for $Z=0.001$, green circles for $Z=0.004$, purple squares for $Z=0.008$, red stars for $Z=0.018$ and magenta upside down triangles for $Z=0.04$. \label{fig11}}
\end{figure}

In Fig. \ref{fig11} the N/O ratio is plotted against the He abundances. In the top panel, the abundances predicted by \citet{karakas10} for stars of different initial masses are labeled with numbers indicating the masses used. Different colours and symbols are used to distinguish between each of the three different values of heavy element abundance $Z$. The majority of the observed points appear to lie between the curves for $Z$ = 0.02 and 0.008, so that there is a general agreement between the predicted and observed abundances. However, some exceptions are noted: the PNe JaSt 23 and JaSt 52 show a very low abundance of He, that is not predicted by the models. Probably neutral helium has a important contribution for the total helium abundance of these PNe. As discussed in Section \ref{subsec:elem_abund}, in the case of helium we opted to do not use any ICF, since the relative populations of helium ions depend essentially on the effective temperature of the central star. So that, there is no reliable way to correct for neutral helium for these objects and the uncertainties are incorporated in the error bars for He abundances. As noted by \citet{garcia-rojas16}, models by \citet{karakas10} cannot predict the low N/O ratio (log(N/O) $< -1.0$) shown by some PNe. They pointed out two possible origins for this discrepancy. In the first one, the initial masses of the progenitor stars for these PNe should be lower than 1 M$_{\sun}$. However this is unlikely since, using the Padova isochrones as revised by \citet{molla09}, the mean life-time for a star with 0.8  M$_{\sun}$ at solar metallicity is 15.8 Gyr, and therefore higher than the expected age of the Universe. The second possibility is that stellar evolution models are predicting too large yields for N. 

Since stellar nucleosynthesis models depend on the micro and macrophysics adopted, it is also interesting to compare the data with different models in the literature. The \citeauthor{ventura14a} group provide a new generation of AGB stellar models that include dust formation in the stellar winds. During the core H-burning phase, the assumptions concern the overshoot of the convective core lead to a less efficient dredge-up and to a lower threshold mass for the activation of the HBB then previous models in the literature. A detailed description of these models are given in \citeauthor{ventura14a} (2014a), for $Z = 0.004$, in \citet{ventura13} for $Z = 0.001$ and $M > 3$ M$_{\sun}$, and in \citeauthor{ventura14b} (2014b) for $Z = 0.001$ and $M < 3 $ M$_{\sun}$. For $Z=0.018$ and $Z=0.04$ the data were obtained from F. Dell'Agli  and J. Garcia-Rojas (private communication). Hereafter we will call all these models as \citeauthor{ventura14a} group. \citet{miller16} provide AGB nucleosynthesis models including an updated treatment of the microphysics (radiative opacities and nuclear reaction rates) and description of the mixing processes and mass loss rates that play a key role during the thermal pulses on the AGB phase. As a result, the new models lead to the occurrence of third dredge up for lower stellar masses. In Fig. \ref{fig11} the results are compared with models from \citet{miller16} (middle panel) and  \citeauthor{ventura14a} group (bottom panel). In this figure, we can observe a fair agreement between the data and the models. The different micro and macrophysics adopted by each model change the progenitor masses at higher N/O ratios. The higher N/O $\sim 0.6$ from our data are compatible with progenitor masses of 3--4 M$_{\sun}$ in the case of the models from \citet{miller16} and of 5--6 M$_{\sun}$ in the case of the models from \citet{karakas10} and \citeauthor{ventura14a} group. Notice that JaSt 52 has very low abundances of O and a lower N/O ratio compared with the other PNe in our sample. Since neutral helium has an important contribution for this PN, the abundances for this object should be taken with caution. 

\begin{figure*}
\includegraphics[width=17cm]{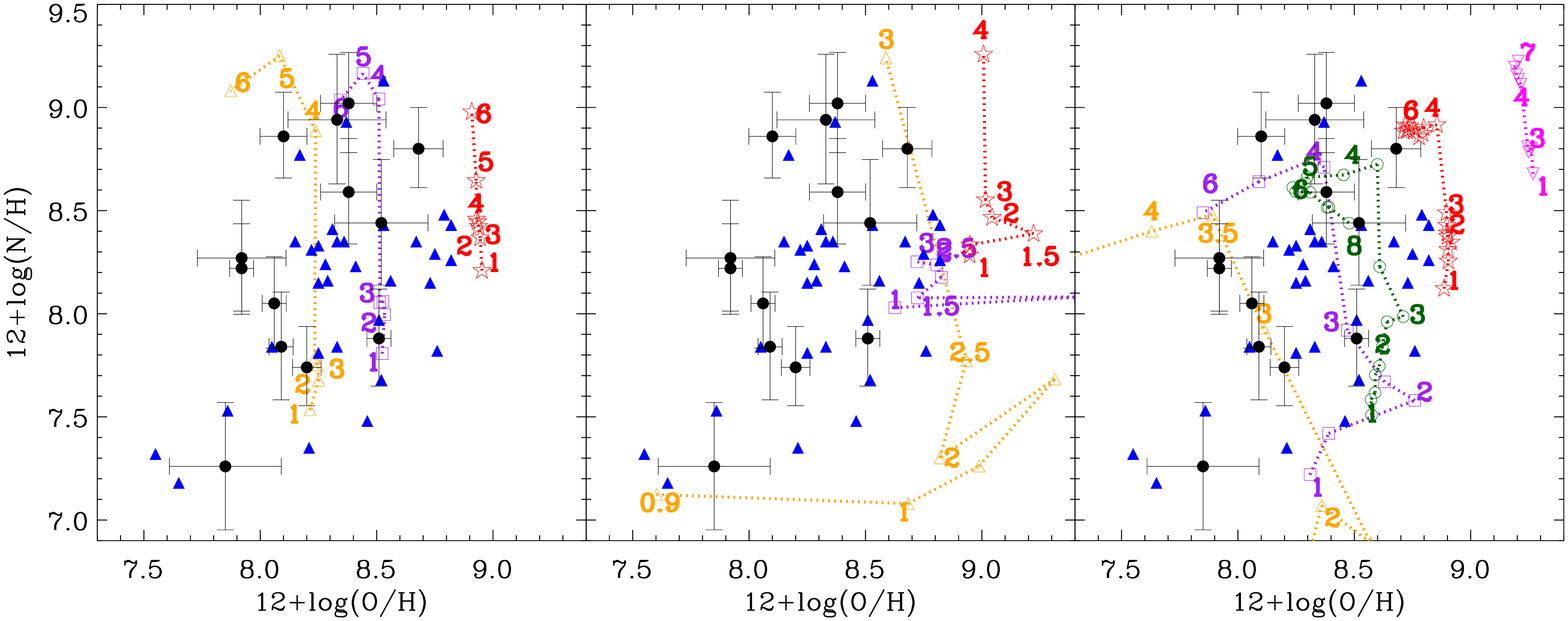}
\caption{$12+\log(\mbox{N/H})$ as a function of $12+\log(\mbox{O/H})$. Left: comparison with the models from \citet{karakas10}; middle: comparison with the models from \citet{miller16}; and right: comparison with the models from \citeauthor{ventura14a} group.  Symbols are as in Fig. \ref{fig11}. \label{fig12}}
\end{figure*}

The N/H abundances as a function of O/H abundances are displayed in Fig. \ref{fig12}. Most of our data is in agreement with the models by \citet{karakas10} for higher metallicities. Exceptions are PNe JaSt 17 and JaSt 52 again (see discussion above) with a low O/H not predicted by the models at the given metallicities. In this figure, the predictions of the models are as expected: more massive stars produce larger amounts of N. In the case of the models by \citet{miller16}, there is an offset towards higher O/H abundances compared with the data. The PNe in our sample PPAJ1740-2708, PPAJ1738-2800, PPAJ1734-3004, JaSt 36 and JaSt 46 have N/H abundances compatible with models for initial stellar masses higher than 4 M$_{\sun}$.  

An important difference between our sample and CCM10 is observed in figures \ref{fig11} and \ref{fig12}: in CCM10 some points are compatible with the lower metallicity model ($Z=0.004$) by \citet{karakas10} and also lower initial masses ($< 4 \mbox{M}_{\sun}$). In our sample a large fraction of PNe have abundances compatible with models at higher metallicities. Also, the superior limit for the initial masses depend on the model adopted. Considering  the results from \citet{miller16}, the PNe near the GC are compatible with stellar initial masses $< 3 \mbox{M}_{\sun}$ for $Z=0.001$. In the case of the models from \citeauthor{ventura14a} group for higher metallicities the data are compatible with initial masses $< 6 \mbox{M}_{\sun}$. On the other hand, \citet{gesicki14} using high resolution imaging and spectroscopic observations of 31 compact PNe derived their central star masses.  Post-AGB evolutionary models were used to fit the white dwarf mass distribution and initial-final mass relations were derived using white dwarfs in clusters. They obtained a mass distribution for GBPNe and  find a mass limit of 2.5~M$_{\sun}$, which is lower than the masses obtained from the models in Fig. \ref{fig11}. However, we note that in \citet{gesicki14} the PNe distribution in their Fig. 9 shows they explored a different region in the bulge since there are no PNe within 2\degr from the GC. 

\begin{figure}
\includegraphics[width=8cm]{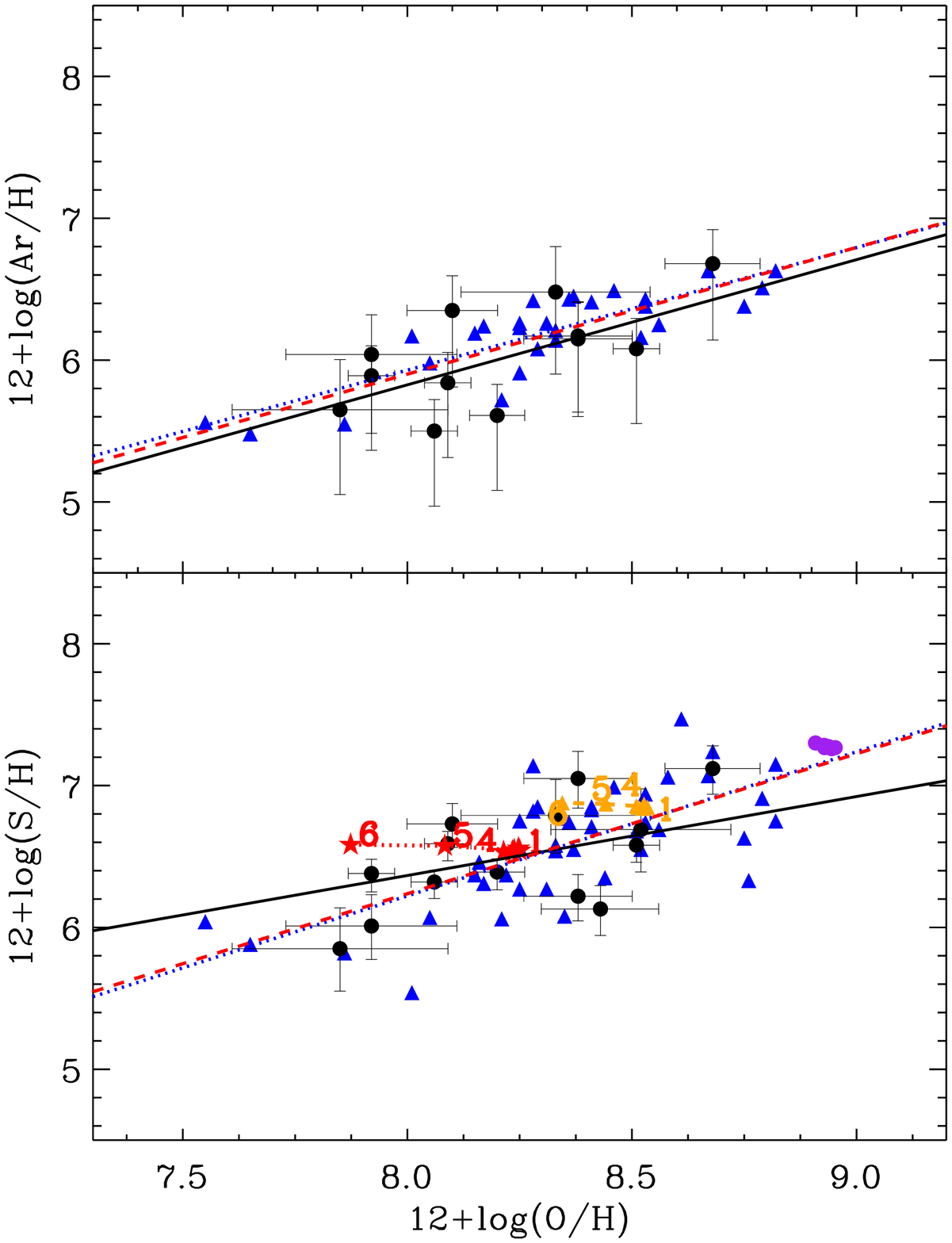}
\caption{Top: $12+\log(\mbox{Ar/H})$ vs. $12+\log(\mbox{O/H})$. Bottom: $12+\log(\mbox{S/H})$ vs. $12+\log(\mbox{O/H})$. Symbols are as in the top panel of Fig. \ref{fig11} for \citet{karakas10} models. The lines are linear fits to the data: only PNe near the GC (black continuous line), only PNe from CCM10 (blue dotted line) and all the data (red dashed line).\label{fig13}}
\end{figure}

The behaviour of Ar/H vs. O/H and S/H vs. O/H are presented Fig. \ref{fig13}.
 A correlation is found for these elemental abundances for both samples.
The linear Pearson correlation coefficients for CCM10 sample are 0.84 and 0.70 for Ar and S, respectively. The slopes of the linear fit are 0.86 and 1.02 for Ar and S,
respectively. Considering only the GC sample, the correlation coefficients
are 0.65 and 0.63 and the slopes 0.88 and 0.56, respectively. The low number of data points and also the uncertainties in the GC sample may wipe out the expected relations. However, considering all the data (this work + CCM10) the linear correlation coefficients are 0.79 and 0.69 for Ar and S. In this case the slopes of the linear fit are
0.89 and 0.99. So that a linear relationship there exist considering both
samples. A detailed discussion of the correlations between neon, sulphur, and 
argon abundances with oxygen in photoionized nebulae of the Local Group is given by \citet{maciel17}. The predictions of evolution models by \citet{karakas10} at Z = 0.004, 0.008 and 0.02 are displayed in the graph for S/H vs. O/H. Clearly, models show that S is not modified by stellar nucleosynthesis, independently of the initial stellar mass. On the contrary, O is expected to be modified in progenitor stars heavier than 4 M$_{\sun}$ at low-metallicity environments.  In the case of Ar, the models by \citet{karakas10} do not give predictions for these elements, so that the abundances could not be compared with theoretical results. 

\begin{figure*}
\includegraphics[width=6cm]{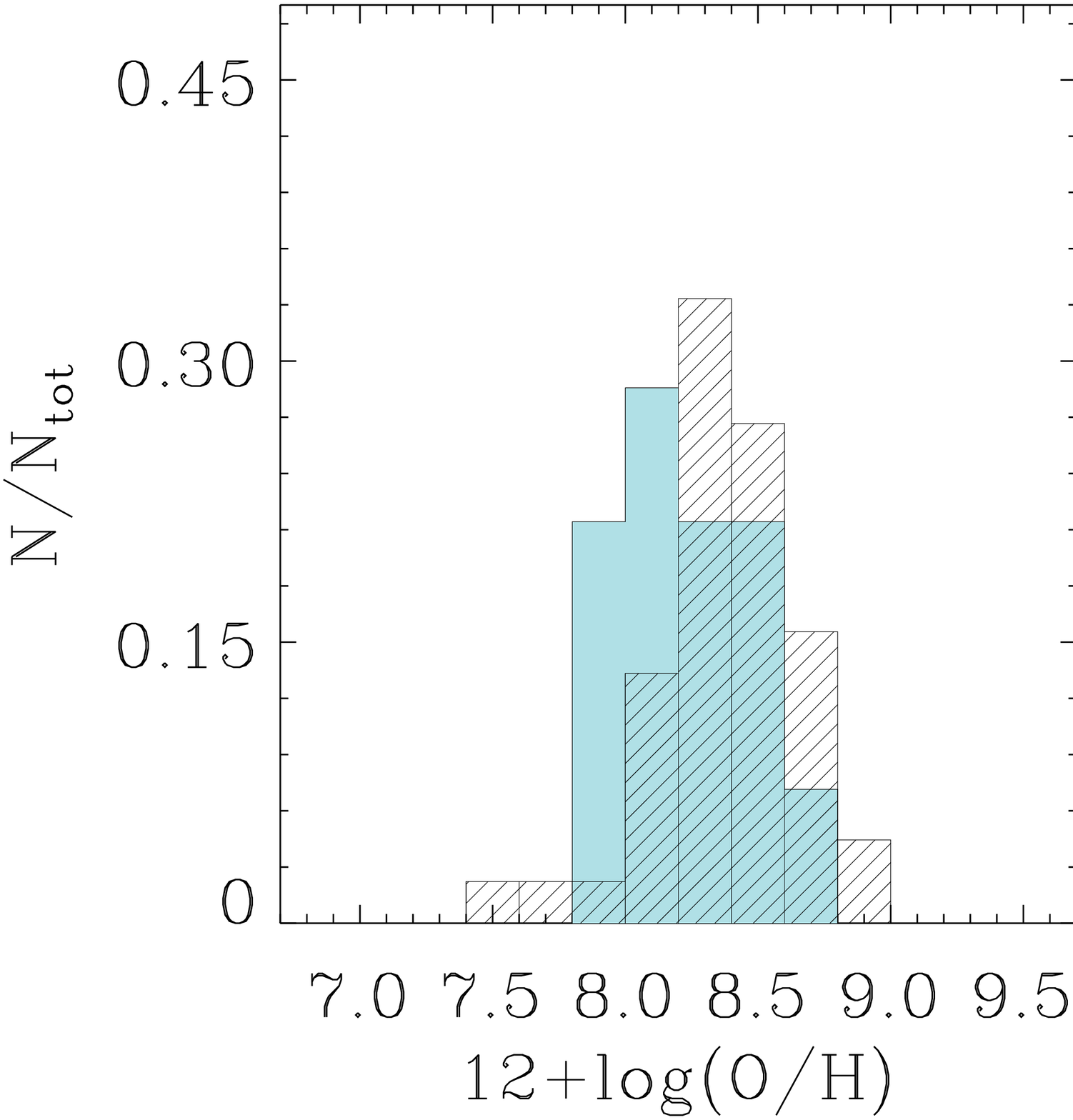}
\includegraphics[width=6cm]{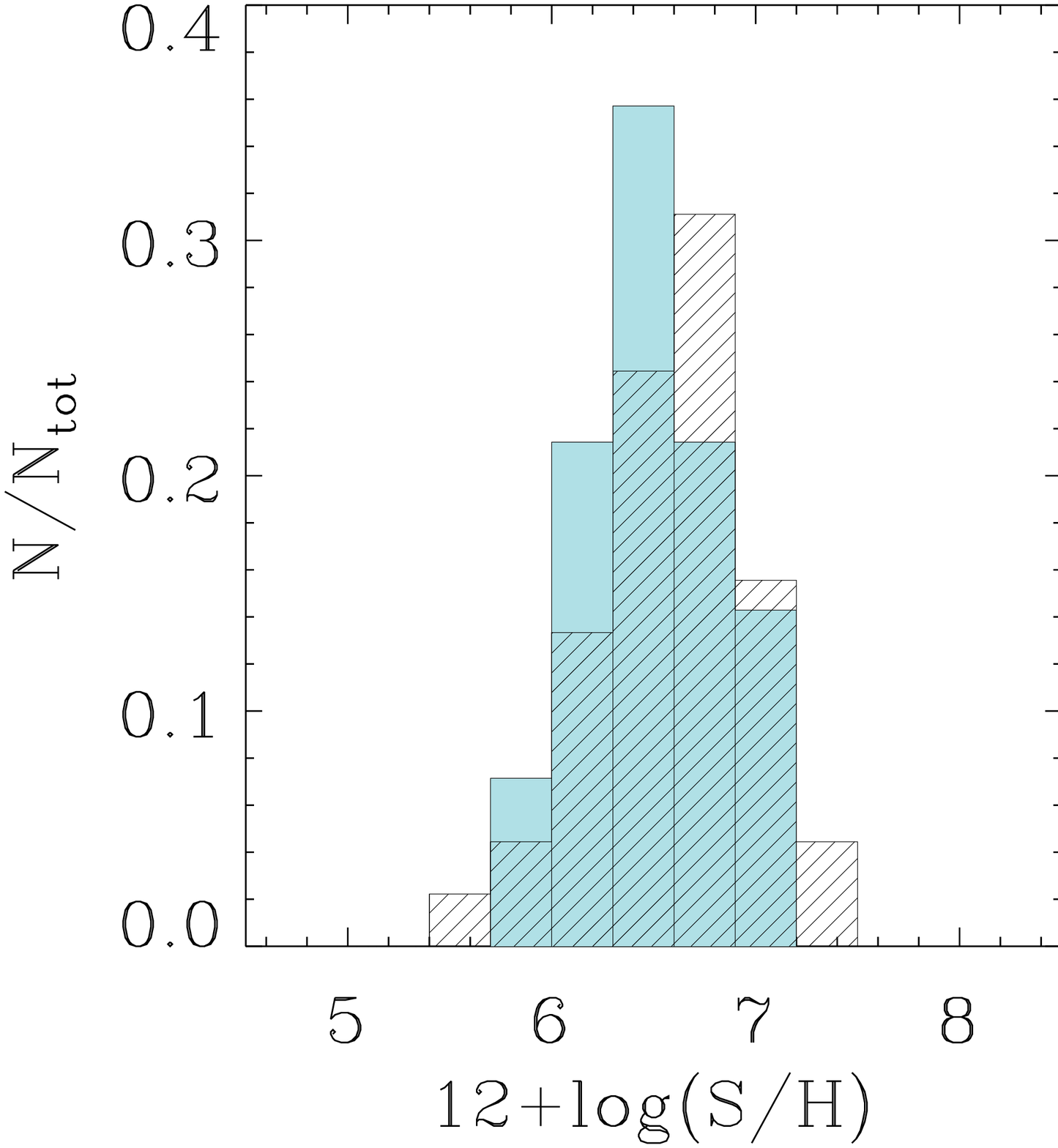}
\includegraphics[width=6cm]{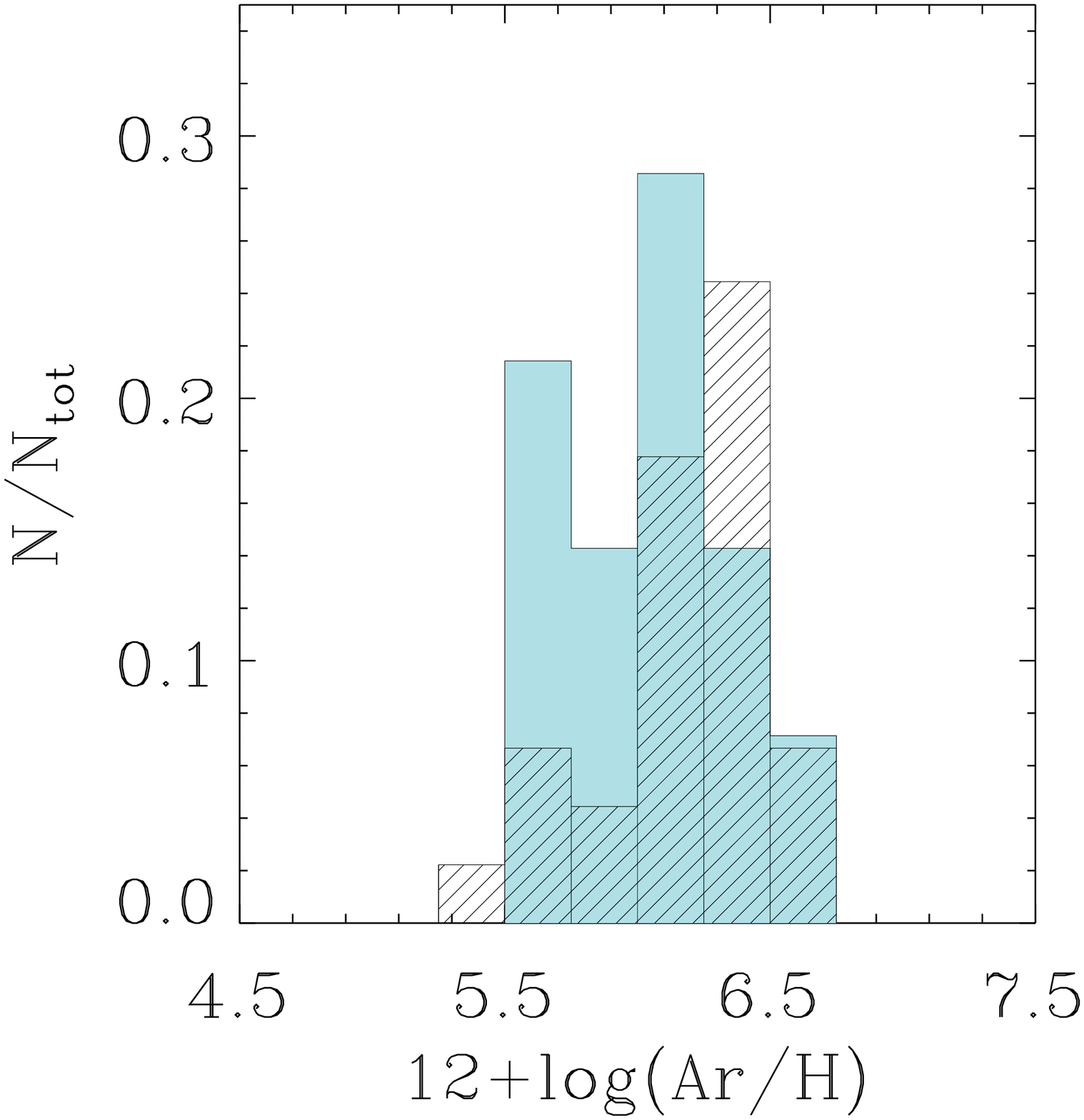}
\includegraphics[width=6cm]{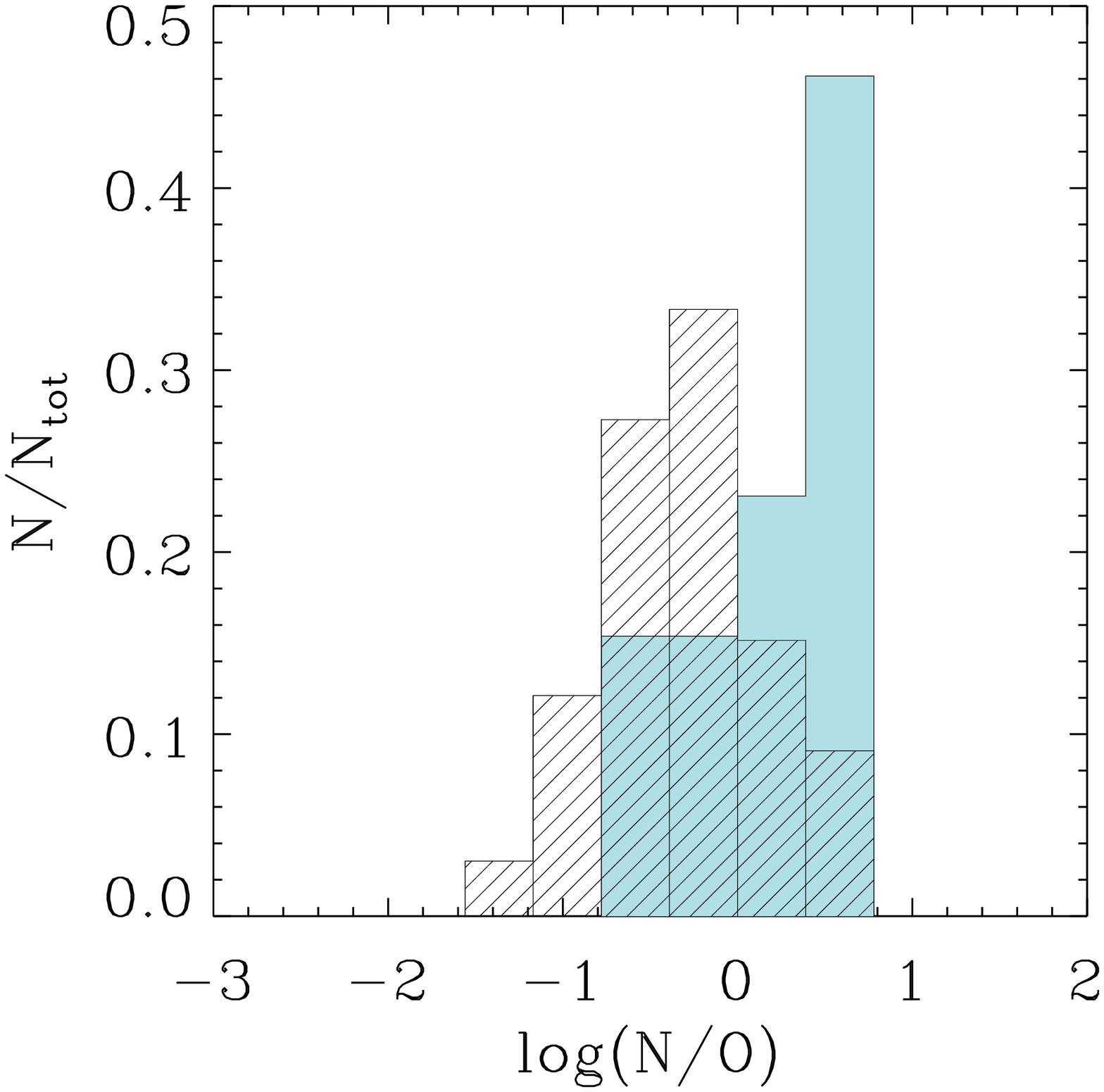}

\caption{Abundances distributions (histograms) for $12+\log(\mbox{O/H})$ (top left), $12+\log(\mbox{S/H})$ (top right), $12+\log(\mbox{Ar/H})$ (bottom left) and $\log(\mbox{N/O})$ (bottom right). Filled histograms are the data from GBPNe near the GC and unfilled histograms represent the that from CCM10 (PNe from outer regions of the bulge). \label{fig14}}
\end{figure*}

The histograms of the abundances distributions for O/H, S/H, A/Hr and $\log(\mbox{N/O})$ are shown in Fig. \ref{fig14}. We do not find important differences between the distributions of  O/H, S/H and Ar/H comparing both samples. The average GC abundances of O/H, S/H, and Ar/H are 0.13, 0.11, and 0.16 dex lower than the average values of the outer bulge sample. However, these differences are within the expected errors in the abundances. On the other hand, despite of the low number of PNe in our sample, some important differences are evident in the histograms of  $\log(\mbox{N/O})$: the distribution for PNe near the GC is shifted for higher values compared with those from CCM10 data. The mean $\log(\mbox{N/O})$ are -0.28 dex and  0.28 dex for CCM10 and GC samples, respectively. So that the difference between both samples in $\log(\mbox{N/O})$ is considerable ($\sim$0.56 dex). Only for Ne the comparison cannot be made since, due to the high interstellar extinction in our spectra, we could not observe the  [Ne\,{\sc iii}] $\lambda$3869 \AA \ and  [Ne\,{\sc iii}] $\lambda$3967 \AA \ lines necessary to calculate the Ne elemental abundances. However, as alerted before,  this result should be interpreted with some caution, since due to the high interstellar extinction in the direction of the GC it is very difficult to define an metallicity-unbiased sample.

It is important to compare the results obtained in this paper with those from stars located near the GC. \citet{ryde16} have measured abundances of Mg, Si, Ca and Fe in 28 M-type giants in the GC region using high-resolution IR spectroscopy. Their data show a trend where the metallicity of the stars increases progressively as closer to the GC the stars are. By means of high-resolution IR spectra, \citet{cunha07} observed a sample of cool stars within 30 pc from the GC. They obtained that  [O/Fe] and [Ca/Fe] are enhanced by 0.2 and 0.3 dex, respectively, relative the Sun Fe/H abundances. Their results pointed that the width of the [Fe/H] distribution in the GC is narrower than the one obtained for the older bulge population (outer bulge population). We note, however, that the data from \citet{cunha07} are located nearer the Galactic centre than the PNe from our sample and the interpretation for the $\alpha$-enhancement found in their sample might be different. Their interpretation is that the $\alpha$-enhancement observed in cool stars near the GC might be due to a IMF weighted toward more massive stars or recent local SN II chemical enrichment within the central 50 pc of the Galaxy, or a mixture of bulge red-giant winds feeding the GC ISM. Regarding the results from other spiral galaxies, \citet{florido15} used a sample of nearby face-on disc galaxies with available SDSS spectra to derive chemical abundances of the ionized gas. Their results point to an enhancement of N/O in barred galaxies compared with non-barred galaxies. Nonetheless, they do not find any difference regarding O/H. This difference in N/O and not in O/H in the centres of barred galaxies could be due to a different star formation efficiency in the inner parts of galaxies as a consequence of the influence of gas flows induced by the bars. Indeed, by means of a chemical evolution model, \citet{cavichia14} have simulated the gas flows induced by the Galactic bar and the influence on the Galactic bulge abundance distributions, finding no important differences. However, the SFR is enhanced in the bulge when there is a radial gas flow towards the centre of the Galaxy. Nevertheless, the N/O distribution was not investigated in that work. In the present work, we do not find evidence for important differences in the  abundances of O/H, S/H and Ar/H in the GC PNe compared with those from the outer regions of the bulge. However, we do find evidence for higher N/O ratios in the GC sample, which is expected for more massive progenitor stars. The results of Fig. \ref{fig11}, \ref{fig12} and \ref{fig14} point to recent SFR occurring in the GC, compared with the outer regions of the Galactic bulge. However, this result should be interpreted with some caution, since due to the high interstellar extinction in the direction of the GC it is very difficult to define a metallicity-unbiased sample.

\section{Conclusions}
\label{sec:conc}

We have performed spectrophotometric observations with the 4.1 m SOAR (Chile) and the 1.6 m OPD/LNA (Brazil) telescopes to obtain physical parameters and chemical abundances for a sample of 17 planetary nebulae located within 2\degr of the Galactic centre. We derived chemical abundances for He, N, O, S and Ar. The results point to high obscured PNe, with E(B-V) roughly 2.3 on the average. With such high extinction, no lines are seen in the blue part of the spectra, at wavelengths shorter than H$\beta$. We have implemented the new ICFs from \citet{delgado-inglada14} for the elemental abundances determination and also the new He\,{\sc i} emissivities from \citet{porter13}. S abundances were derived using optical and NIR lines, reducing the uncertainties associated with S ICFs. The abundances predicted by \citet{karakas10}, \citet{miller16} and \citeauthor{ventura13} group for stars of different initial masses and metallicty were used to constraint the masses and initial metallicity of the progenitor stars. An important difference between our sample near the Galactic centre and PNe located in the outer parts of the bulge is observed. In our previous work \citep[][outer bulge region]{cavichia10} some points are compatible with the lower metallicity model ($Z=0.004$) by \citet{karakas10} and also lower initial masses ($< 4 \mbox{M}_{\sun}$). In the PNe located near the Galactic centre, a large fraction of PNe have abundances compatible with models at higher progenitor masses. The results point to a considerable difference in $\log(\mbox{N/O})$, the PNe near the GC enhanced on average by $\sim$0.56 dex  compared with PNe in the outer regions of the bulge. A higher N/O ratio is expected for more massive progenitor stars and may indicate recent episodes of star formation taking place at the GC, compared with the outer regions of the Galactic bulge.

A large percentage of PNe originated from stars formed in a high-metallicity environment is not expected in a bulge that originates purely from a gravitational collapse or by hierarchical merging of smaller objects. In this case, the formation process is generally fast and occurs earlier in the Galaxy formation process, before the present disc was formed. On the other hand, in bulges formed via disc instabilities (box/peanut bulges), the material that form the bulge is originated by the rearrangement of the disc material through the secular evolution. Since the secular evolution causes a significant amount of star formation within the centre of galaxies \citep{ellison11}, the stellar populations near the central parsecs of these galaxies are rejuvenating \citep{coelho11}. In this scenario, a significant population of young stars is expected. However, a composite scenario with the possibility of a small classical bulge embedded within the box/peanut is not ruled out. The presence of multiple metallicity distribution within a bulge was noted in dissipative collapse models \citep[e.g.][]{samland03} and also in bulges from cosmological simulations \citep{obreja13}. Nonetheless, recently \citet{ness14} demonstrated that old stars are not exclusively linked to a classical bulge and the presence of young stars that are located close to the plane is expected for a bulge that has formed from the disc via dynamical instabilities. In summary, the results found in this paper collaborate to understand the chemical enrichment occurring at the Milky Way central regions. In spite of the results found, it is very difficult to define a complete sample in the GC region. Therefore, due to the low number of PNe and possible selection effects, more data of the same region as presented in this paper are necessary to draw firm conclusions about the chemical enrichment at the inner 2\degr of the Galactic bulge.

\section*{Acknowledgments}
   
This work has been financially supported by the Brazilian agency FAPESP (Proc. 2007/07704-2, 2010/18835-3 and 2012/01017-1). We are grateful to the referee Dr. Zijlstra for his comments and careful reading of the manuscript. O.C. would like to thank J. Garcia-Rojas and F. Dell'Agli for kindly provide the data from AGB nucleosynthesis models. 

\bibliographystyle{mn2e}

\bibliography{biblio}

\begin{thebibliography}{}

\bibitem[\protect\citeauthoryear{{Aver}, {Olive} \& {Skillman}}{{Aver}
  et~al.}{2010}]{aver10}
{Aver} E.,  {Olive} K.~A.,    {Skillman} E.~D.,  2010, \jcap, 5, 3

\bibitem[\protect\citeauthoryear{{Bland-Hawthorn} \&
  {Gerhard}}{{Bland-Hawthorn} \& {Gerhard}}{2016}]{bland-hawthorn16}
{Bland-Hawthorn} J.,  {Gerhard} O.,  2016, \araa, 54, 529

\bibitem[\protect\citeauthoryear{{Cahn}, {Kaler} \& {Stanghellini}}{{Cahn}
  et~al.}{1992}]{cahn92}
{Cahn} J.~H.,  {Kaler} J.~B.,    {Stanghellini} L.,  1992, \aaps, 94, 399

\bibitem[\protect\citeauthoryear{{Cardelli}, {Clayton} \& {Mathis}}{{Cardelli}
  et~al.}{1989}]{cardelli89}
{Cardelli} J.~A.,  {Clayton} G.~C.,    {Mathis} J.~S.,  1989, \apj, 345, 245

\bibitem[\protect\citeauthoryear{{Cavichia}, {Costa} \& {Maciel}}{{Cavichia}
  et~al.}{2010}]{cavichia10}
{Cavichia} O.,  {Costa} R.~D.~D.,    {Maciel} W.~J.,  2010, RMxAA, 46,
  159(CCM10)

\bibitem[\protect\citeauthoryear{{Cavichia}, {Costa} \& {Maciel}}{{Cavichia}
  et~al.}{2011}]{cavichia11}
{Cavichia} O.,  {Costa} R.~D.~D.,    {Maciel} W.~J.,  2011, RMxAA, 47, 49

\bibitem[\protect\citeauthoryear{{Cavichia}, {Moll{\'a}}, {Costa} \&
  {Maciel}}{{Cavichia} et~al.}{2014}]{cavichia14}
{Cavichia} O.,  {Moll{\'a}} M.,  {Costa} R.~D.~D.,    {Maciel} W.~J.,  2014,
  \mnras, 437, 3688

\bibitem[\protect\citeauthoryear{{Chiappini}, {G{\'o}rny}, {Stasi{\'n}ska} \&
  {Barbuy}}{{Chiappini} et~al.}{2009}]{chiappini09}
{Chiappini} C.,  {G{\'o}rny} S.~K.,  {Stasi{\'n}ska} G.,    {Barbuy} B.,  2009,
  \aap, 494, 591

\bibitem[\protect\citeauthoryear{{Clayton}}{{Clayton}}{1968}]{clayton68}
{Clayton} D.~D.,  1968, {Principles of stellar evolution and nucleosynthesis}.
New York: McGraw-Hill, 1968

\bibitem[\protect\citeauthoryear{{Coelho} \& {Gadotti}}{{Coelho} \&
  {Gadotti}}{2011}]{coelho11}
{Coelho} P.,  {Gadotti} D.~A.,  2011, \apjl, 743, L13

\bibitem[\protect\citeauthoryear{{Cuisinier}, {Maciel}, {K{\"o}ppen}, {Acker}
  \& {Stenholm}}{{Cuisinier} et~al.}{2000}]{cuisinier00}
{Cuisinier} F.,  {Maciel} W.~J.,  {K{\"o}ppen} J.,  {Acker} A.,    {Stenholm}
  B.,  2000, \aap, 353, 543

\bibitem[\protect\citeauthoryear{{Cunha}, {Sellgren}, {Smith}, {Ramirez},
  {Blum} \& {Terndrup}}{{Cunha} et~al.}{2007}]{cunha07}
{Cunha} K.,  {Sellgren} K.,  {Smith} V.~V.,  {Ramirez} S.~V.,  {Blum} R.~D.,
  {Terndrup} D.~M.,  2007, \apj, 669, 1011

\bibitem[\protect\citeauthoryear{{Delgado-Inglada}, {Morisset} \&
  {Stasi{\'n}ska}}{{Delgado-Inglada} et~al.}{2014}]{delgado-inglada14}
{Delgado-Inglada} G.,  {Morisset} C.,    {Stasi{\'n}ska} G.,  2014, \mnras,
  440, 536

\bibitem[\protect\citeauthoryear{{Eggen}, {Lynden-Bell} \& {Sandage}}{{Eggen}
  et~al.}{1962}]{eggen62}
{Eggen} O.~J.,  {Lynden-Bell} D.,    {Sandage} A.~R.,  1962, \apj, 136, 748

\bibitem[\protect\citeauthoryear{{Ellison}, {Nair}, {Patton}, {Scudder},
  {Mendel} \& {Simard}}{{Ellison} et~al.}{2011}]{ellison11}
{Ellison} S.~L.,  {Nair} P.,  {Patton} D.~R.,  {Scudder} J.~M.,  {Mendel}
  J.~T.,    {Simard} L.,  2011, \mnras, 416, 2182

\bibitem[\protect\citeauthoryear{{Escudero} \& {Costa}}{{Escudero} \&
  {Costa}}{2001}]{escudero01}
{Escudero} A.~V.,  {Costa} R.~D.~D.,  2001, \aap, 380, 300

\bibitem[\protect\citeauthoryear{{Escudero}, {Costa} \& {Maciel}}{{Escudero}
  et~al.}{2004}]{escudero04}
{Escudero} A.~V.,  {Costa} R.~D.~D.,    {Maciel} W.~J.,  2004, \aap, 414, 211

\bibitem[\protect\citeauthoryear{{Exter}, {Barlow} \& {Walton}}{{Exter}
  et~al.}{2004}]{exter04}
{Exter} K.~M.,  {Barlow} M.~J.,    {Walton} N.~A.,  2004, \mnras, 349, 1291

\bibitem[\protect\citeauthoryear{{Fitzpatrick}}{{Fitzpatrick}}{1999}]{fitzpatr%
ick99}
{Fitzpatrick} E.~L.,  1999, \pasp, 111, 63

\bibitem[\protect\citeauthoryear{Fitzpatrick \& Massa}{Fitzpatrick \&
  Massa}{2009}]{fitzpatrick09}
Fitzpatrick E.~L.,  Massa D.,  2009, \apj, 699, 1209

\bibitem[\protect\citeauthoryear{{Florido}, {Zurita}, {P{\'e}rez},
  {P{\'e}rez-Montero}, {Coelho} \& {Gadotti}}{{Florido}
  et~al.}{2015}]{florido15}
{Florido} E.,  {Zurita} A.,  {P{\'e}rez} I.,  {P{\'e}rez-Montero} E.,  {Coelho}
  P.~R.~T.,    {Gadotti} D.~A.,  2015, \aap, 584, A88

\bibitem[\protect\citeauthoryear{{Frew} \& {Parker}}{{Frew} \&
  {Parker}}{2010}]{frew10}
{Frew} D.~J.,  {Parker} Q.~A.,  2010, \pasa, 27, 129

\bibitem[\protect\citeauthoryear{{Garc{\'{\i}}a-Rojas}, {Pe{\~n}a},
  {Flores-Dur{\'a}n} \& {Hern{\'a}ndez-Mart{\'{\i}}nez}}{{Garc{\'{\i}}a-Rojas}
  et~al.}{2016}]{garcia-rojas16}
{Garc{\'{\i}}a-Rojas} J.,  {Pe{\~n}a} M.,  {Flores-Dur{\'a}n} S.,
  {Hern{\'a}ndez-Mart{\'{\i}}nez} L.,  2016, \aap, 586, A59

\bibitem[\protect\citeauthoryear{{Gesicki}, {Zijlstra}, {Hajduk} \&
  {Szyszka}}{{Gesicki} et~al.}{2014}]{gesicki14}
{Gesicki} K.,  {Zijlstra} A.~A.,  {Hajduk} M.,    {Szyszka} C.,  2014, \aap,
  566, A48

\bibitem[\protect\citeauthoryear{{Gonzalez}, {Rejkuba}, {Zoccali}, {Valenti},
  {Minniti}, {Schultheis}, {Tobar} \& {Chen}}{{Gonzalez}
  et~al.}{2012}]{gonzalez12}
{Gonzalez} O.~A.,  {Rejkuba} M.,  {Zoccali} M.,  {Valenti} E.,  {Minniti} D.,
  {Schultheis} M.,  {Tobar} R.,    {Chen} B.,  2012, \aap, 543, A13

\bibitem[\protect\citeauthoryear{{G{\'o}rny}, {Chiappini}, {Stasi{\'n}ska} \&
  {Cuisinier}}{{G{\'o}rny} et~al.}{2009}]{gorny09}
{G{\'o}rny} S.~K.,  {Chiappini} C.,  {Stasi{\'n}ska} G.,    {Cuisinier} F.,
  2009, \aap, 500, 1089

\bibitem[\protect\citeauthoryear{{Hamuy}, {Suntzeff}, {Heathcote}, {Walker},
  {Gigoux} \& {Phillips}}{{Hamuy} et~al.}{1994}]{hamuy94}
{Hamuy} M.,  {Suntzeff} N.~B.,  {Heathcote} S.~R.,  {Walker} A.~R.,  {Gigoux}
  P.,    {Phillips} M.~M.,  1994, \pasp, 106, 566

\bibitem[\protect\citeauthoryear{{Hamuy}, {Walker}, {Suntzeff}, {Gigoux},
  {Heathcote} \& {Phillips}}{{Hamuy} et~al.}{1992}]{hamuy92}
{Hamuy} M.,  {Walker} A.~R.,  {Suntzeff} N.~B.,  {Gigoux} P.,  {Heathcote}
  S.~R.,    {Phillips} M.~M.,  1992, \pasp, 104, 533

\bibitem[\protect\citeauthoryear{{Henry}, {Kwitter} \& {Balick}}{{Henry}
  et~al.}{2004}]{henry04}
{Henry} R.~B.~C.,  {Kwitter} K.~B.,    {Balick} B.,  2004, \aj, 127, 2284

\bibitem[\protect\citeauthoryear{{Henry}, {Kwitter}, {Jaskot}, {Balick},
  {Morrison} \& {Milingo}}{{Henry} et~al.}{2010}]{henry10}
{Henry} R.~B.~C.,  {Kwitter} K.~B.,  {Jaskot} A.~E.,  {Balick} B.,  {Morrison}
  M.~A.,    {Milingo} J.~B.,  2010, \apj, 724, 748

\bibitem[\protect\citeauthoryear{{Henry}, {Speck}, {Karakas}, {Ferland} \&
  {Maguire}}{{Henry} et~al.}{2012}]{henry12}
{Henry} R.~B.~C.,  {Speck} A.,  {Karakas} A.~I.,  {Ferland} G.~J.,    {Maguire}
  M.,  2012, \apj, 749, 61

\bibitem[\protect\citeauthoryear{{Izotov}, {Stasi{\'n}ska}, {Meynet}, {Guseva}
  \& {Thuan}}{{Izotov} et~al.}{2006}]{izotov06}
{Izotov} Y.~I.,  {Stasi{\'n}ska} G.,  {Meynet} G.,  {Guseva} N.~G.,    {Thuan}
  T.~X.,  2006, \aap, 448, 955

\bibitem[\protect\citeauthoryear{{Jacoby} \& {Van de Steene}}{{Jacoby} \& {Van
  de Steene}}{2004}]{jacoby04}
{Jacoby} G.~H.,  {Van de Steene} G.,  2004, \aap, 419, 563(JS04)

\bibitem[\protect\citeauthoryear{{Karakas}}{{Karakas}}{2010}]{karakas10}
{Karakas} A.~I.,  2010, \mnras, 403, 1413

\bibitem[\protect\citeauthoryear{{Kingsburgh} \& {Barlow}}{{Kingsburgh} \&
  {Barlow}}{1994}]{kingsburgh94}
{Kingsburgh} R.~L.,  {Barlow} M.~J.,  1994, \mnras, 271, 257

\bibitem[\protect\citeauthoryear{{Maciel}, {Costa} \& {Cavichia}}{{Maciel}
  et~al.}{2017}]{maciel17}
{Maciel} W.~J.,  {Costa} R.~D.~D.,    {Cavichia} O.,  2017, RMxAA, submitted

\bibitem[\protect\citeauthoryear{{Maciel}, {Lago} \& {Costa}}{{Maciel}
  et~al.}{2006}]{maciel06}
{Maciel} W.~J.,  {Lago} L.~G.,    {Costa} R.~D.~D.,  2006, \aap, 453, 587

\bibitem[\protect\citeauthoryear{{Malkin}}{{Malkin}}{2013}]{malkin13}
{Malkin} Z.,  2013, in {de Grijs} R.,  ed., IAU Symposium Vol.~289 of IAU
  Symposium, {Statistical analysis of the determinations of the Sun's
  Galactocentric distance}.
pp 406--409

\bibitem[\protect\citeauthoryear{{Milingo}, {Kwitter}, {Henry} \&
  {Souza}}{{Milingo} et~al.}{2010}]{milingo10}
{Milingo} J.~B.,  {Kwitter} K.~B.,  {Henry} R.~B.~C.,    {Souza} S.~P.,  2010,
  \apj, 711, 619

\bibitem[\protect\citeauthoryear{{Miller Bertolami}}{{Miller
  Bertolami}}{2016}]{miller16}
{Miller Bertolami} M.~M.,  2016, \aap, 588, A25

\bibitem[\protect\citeauthoryear{{Miszalski}, {Acker}, {Moffat}, {Parker} \&
  {Udalski}}{{Miszalski} et~al.}{2009}]{miszalski09}
{Miszalski} B.,  {Acker} A.,  {Moffat} A.~F.~J.,  {Parker} Q.~A.,    {Udalski}
  A.,  2009, \aap, 496, 813

\bibitem[\protect\citeauthoryear{{Miszalski}, {Parker}, {Acker}, {Birkby},
  {Frew} \& {Kovacevic}}{{Miszalski} et~al.}{2008}]{miszalski08}
{Miszalski} B.,  {Parker} Q.~A.,  {Acker} A.,  {Birkby} J.~L.,  {Frew} D.~J.,
   {Kovacevic} A.,  2008, \mnras, 384, 525

\bibitem[\protect\citeauthoryear{{Moll{\'a}}, {Garc{\'{\i}}a-Vargas} \&
  {Bressan}}{{Moll{\'a}} et~al.}{2009}]{molla09}
{Moll{\'a}} M.,  {Garc{\'{\i}}a-Vargas} M.~L.,    {Bressan} A.,  2009, \mnras,
  398, 451

\bibitem[\protect\citeauthoryear{{Monreal-Ibero}, {Walsh}, {Westmoquette} \&
  {V{\'{\i}}lchez}}{{Monreal-Ibero} et~al.}{2013}]{monreal-ibero13}
{Monreal-Ibero} A.,  {Walsh} J.~R.,  {Westmoquette} M.~S.,    {V{\'{\i}}lchez}
  J.~M.,  2013, \aap, 553, A57

\bibitem[\protect\citeauthoryear{{Nataf}, {Gonzalez}, {Casagrande} \& {et
  al.}}{{Nataf} et~al.}{2016}]{nataf16}
{Nataf} D.~M.,  {Gonzalez} O.~A.,  {Casagrande} L.,    {et al.} 2016, \mnras,
  456, 2692

\bibitem[\protect\citeauthoryear{{Nataf}, {Gould}, {Fouqu{\'e}} \& {et
  al.}}{{Nataf} et~al.}{2013}]{nataf13}
{Nataf} D.~M.,  {Gould} A.,  {Fouqu{\'e}} P.,    {et al.} 2013, \apj, 769, 88

\bibitem[\protect\citeauthoryear{{Ness}, {Debattista}, {Bensby}, {Feltzing},
  {Ro{\v s}kar}, {Cole}, {Johnson} \& {Freeman}}{{Ness} et~al.}{2014}]{ness14}
{Ness} M.,  {Debattista} V.~P.,  {Bensby} T.,  {Feltzing} S.,  {Ro{\v s}kar}
  R.,  {Cole} D.~R.,  {Johnson} J.~A.,    {Freeman} K.,  2014, \apjl, 787, L19

\bibitem[\protect\citeauthoryear{{Obreja}, {Dom{\'{\i}}nguez-Tenreiro},
  {Brook}, {Mart{\'{\i}}nez-Serrano}, {Dom{\'e}nech-Moral}, {Serna},
  {Moll{\'a}} \& {Stinson}}{{Obreja} et~al.}{2013}]{obreja13}
{Obreja} A.,  {Dom{\'{\i}}nguez-Tenreiro} R.,  {Brook} C.,
  {Mart{\'{\i}}nez-Serrano} F.~J.,  {Dom{\'e}nech-Moral} M.,  {Serna} A.,
  {Moll{\'a}} M.,    {Stinson} G.,  2013, \apj, 763, 26

\bibitem[\protect\citeauthoryear{{Osterbrock} \& {Ferland}}{{Osterbrock} \&
  {Ferland}}{2006}]{osterbrock06}
{Osterbrock} D.~E.,  {Ferland} G.~J.,  2006, {\itshape Astrophysics of gaseous
  nebulae and active galactic nuclei}.
2nd ed.; Sausalito, CA: Univ. Science Books

\bibitem[\protect\citeauthoryear{{Parker}, {Acker}, {Frew} \& {et
  al.}}{{Parker} et~al.}{2006}]{parker06}
{Parker} Q.~A.,  {Acker} A.,  {Frew} D.,    {et al.} 2006, \mnras, 373, 79

\bibitem[\protect\citeauthoryear{{Peyaud}, {Boily}, {Acker} \&
  {Parker}}{{Peyaud} et~al.}{2006}]{peyaud06}
{Peyaud} A.~E.~J.,  {Boily} C.,  {Acker} A.,    {Parker} Q.,  2006, in {Barlow}
  M.~J.,  {M{\'e}ndez} R.~H.,  eds, Planetary Nebulae in our Galaxy and Beyond
  Vol.~234 of IAU Symposium, {\itshape Kinematics and Dynamics of the Galactic
  Bulge through Planetary Nebulae}.
pp 485--486

\bibitem[\protect\citeauthoryear{{Porter}, {Ferland}, {Storey} \&
  {Detisch}}{{Porter} et~al.}{2012}]{porter12}
{Porter} R.~L.,  {Ferland} G.~J.,  {Storey} P.~J.,    {Detisch} M.~J.,  2012,
  \mnras, 425, L28

\bibitem[\protect\citeauthoryear{{Porter}, {Ferland}, {Storey} \&
  {Detisch}}{{Porter} et~al.}{2013}]{porter13}
{Porter} R.~L.,  {Ferland} G.~J.,  {Storey} P.~J.,    {Detisch} M.~J.,  2013,
  \mnras, 433, L89

\bibitem[\protect\citeauthoryear{{Ryde}, {Schultheis}, {Grieco}, {Matteucci},
  {Rich} \& {Uttenthaler}}{{Ryde} et~al.}{2016}]{ryde16}
{Ryde} N.,  {Schultheis} M.,  {Grieco} V.,  {Matteucci} F.,  {Rich} R.~M.,
  {Uttenthaler} S.,  2016, \aj, 151, 1

\bibitem[\protect\citeauthoryear{{Samland} \& {Gerhard}}{{Samland} \&
  {Gerhard}}{2003}]{samland03}
{Samland} M.,  {Gerhard} O.~E.,  2003, \aap, 399, 961

\bibitem[\protect\citeauthoryear{{Shaw} \& {Dufour}}{{Shaw} \&
  {Dufour}}{1995}]{shaw95}
{Shaw} R.~A.,  {Dufour} R.~J.,  1995, \pasp, 107, 896

\bibitem[\protect\citeauthoryear{{Stanghellini}, {Shaw} \&
  {Villaver}}{{Stanghellini} et~al.}{2008}]{stanghellini08}
{Stanghellini} L.,  {Shaw} R.~A.,    {Villaver} E.,  2008, \apj, 689, 194

\bibitem[\protect\citeauthoryear{{Stasi{\'n}ska}, {Richer} \&
  {McCall}}{{Stasi{\'n}ska} et~al.}{1998}]{stasinska98}
{Stasi{\'n}ska} G.,  {Richer} M.~G.,    {McCall} M.~L.,  1998, \aap, 336, 667

\bibitem[\protect\citeauthoryear{{Storey} \& {Zeippen}}{{Storey} \&
  {Zeippen}}{2000}]{storey00}
{Storey} P.~J.,  {Zeippen} C.~J.,  2000, \mnras, 312, 813

\bibitem[\protect\citeauthoryear{{Uscanga}, {G{\'o}mez}, {Su{\'a}rez} \&
  {Miranda}}{{Uscanga} et~al.}{2012}]{uscanga12}
{Uscanga} L.,  {G{\'o}mez} J.~F.,  {Su{\'a}rez} O.,    {Miranda} L.~F.,  2012,
  \aap, 547, A40

\bibitem[\protect\citeauthoryear{{van Dokkum}}{{van
  Dokkum}}{2001}]{vandokkun01}
{van Dokkum} P.~G.,  2001, \pasp, 113, 1420

\bibitem[\protect\citeauthoryear{{Ventura}, {Criscienzo}, {D'Antona},
  {Vesperini}, {Tailo}, {Dell'Agli} \& {D'Ercole}}{{Ventura}
  et~al.}{2014}]{ventura14a}
{Ventura} P.,  {Criscienzo} M.~D.,  {D'Antona} F.,  {Vesperini} E.,  {Tailo}
  M.,  {Dell'Agli} F.,    {D'Ercole} A.,  2014, \mnras, 437, 3274

\bibitem[\protect\citeauthoryear{{Ventura}, {Dell'Agli}, {Schneider}, {Di
  Criscienzo}, {Rossi}, {La Franca}, {Gallerani} \& {Valiante}}{{Ventura}
  et~al.}{2014}]{ventura14b}
{Ventura} P.,  {Dell'Agli} F.,  {Schneider} R.,  {Di Criscienzo} M.,  {Rossi}
  C.,  {La Franca} F.,  {Gallerani} S.,    {Valiante} R.,  2014, \mnras, 439,
  977

\bibitem[\protect\citeauthoryear{{Ventura}, {Di Criscienzo}, {Carini} \&
  {D'Antona}}{{Ventura} et~al.}{2013}]{ventura13}
{Ventura} P.,  {Di Criscienzo} M.,  {Carini} R.,    {D'Antona} F.,  2013,
  \mnras, 431, 3642

\bibitem[\protect\citeauthoryear{{Wang} \& {Liu}}{{Wang} \&
  {Liu}}{2007}]{wang07}
{Wang} W.,  {Liu} X.-W.,  2007, \mnras, 381, 669

\bibitem[\protect\citeauthoryear{{Weiland}, {Arendt}, {Berriman}, {Dwek},
  {Freudenreich}, {Hauser}, {Kelsall}, {Lisse}, {Mitra}, {Moseley}, {Odegard},
  {Silverberg}, {Sodroski}, {Spiesman} \& {Stemwedel}}{{Weiland}
  et~al.}{1994}]{weiland94}
{Weiland} J.~L.,  {Arendt} R.~G.,  {Berriman} G.~B.,  {Dwek} E.,
  {Freudenreich} H.~T.,  {Hauser} M.~G.,  {Kelsall} T.,  {Lisse} C.~M.,
  {Mitra} M.,  {Moseley} S.~H.,  {Odegard} N.~P.,  {Silverberg} R.~F.,
  {Sodroski} T.~J.,  {Spiesman} W.~J.,    {Stemwedel} S.~W.,  1994, \apjl, 425,
  L81

\bibitem[\protect\citeauthoryear{{Zijlstra} \& {Pottasch}}{{Zijlstra} \&
  {Pottasch}}{1991}]{zijlstra91}
{Zijlstra} A.~A.,  {Pottasch} S.~R.,  1991, \aap, 243, 478

\bibitem[\protect\citeauthoryear{{Zinn}}{{Zinn}}{1985}]{zinn85}
{Zinn} R.,  1985, \apj, 293, 424

\end{thebibliography}

\end{document}